\documentclass[useAMS,usenatbib]{mn2e}

\usepackage{graphicx}
\usepackage{amsmath}
\usepackage{amssymb}

\numberwithin{equation}{section}

\title[Dust Settling and Planetary Migration]
{Dust Settling and Rapid Planetary Migration}
\author[Y. Hasegawa and R. E. Pudritz]{Yasuhiro Hasegawa$^{1}$\thanks{E-mail:
hasegay@physics.mcmaster.ca (YH); pudritz@physics.mcmaster.ca (REP)} and Ralph E. Pudritz$^{1,2}$\footnotemark[1]\\
$^{1}$Department of Physics and Astronomy, McMaster University, Hamilton, ON L8S 4M1, Canada\\
$^{2}$Origins Institute, McMaster University, Hamilton, ON L8S 4M1, Canada}

\begin{document}

\date{}

\pagerange{\pageref{firstpage}--\pageref{lastpage}} \pubyear{2009}

\maketitle

\label{firstpage}

\begin{abstract}

Planetary migration is essential to explain the observed mass-period relation for exoplanets. 
Without some stopping mechanism, the tidal, resonant interaction between planets and their gaseous disc 
generally causes the planets to migrate inward so efficiently that they plunge into the host star within 
the gaseous disc lifetime ($\sim $ 1-3 Myrs). We investigate planetary migration by analytically calculating 
the migration rate and time within self-consistently computed, radiatively heated discs around M stars in which 
the effects of dust settling are included. We show that dust settling lowers the disc temperature and raises the 
gas density in the mid-plane. This inescapable evolution of disc structure speeds up type I planetary migration 
for lower mass bodies by up to a factor of about 2. We also examine the effects of dust settling on the gap-opening 
mass and type II migration, and find that the gap-opening mass is reduced by a factor of 2 and type II migration 
becomes slower by a factor of 2. While dust settling can somewhat alleviate the problem of planetary migration 
for more massive planets, the more rapid migration of low mass planets and planetary cores requires a robust 
slowing mechanism.  
    
\end{abstract}

\begin{keywords}
accretion, accretion discs -- radiative transfer -- turbulence -- methods:analytical -- 
planetary systems:protoplanetary discs
\end{keywords}

\section{Introduction}

Planet formation involves the interaction between protoplanets and gaseous protoplanetary discs. This 
interaction drives radial orbital drifts for protoplanets \citep[hereafter GT80]{gt79,gt80}, possibly 
resulting in the diversity of exoplanetary orbital properties captured in the observed mass-period 
relation \citep{us07}. In standard disc models, planets lose their net angular momentum by resonant interaction 
with their natal discs and migrate inward. This is confirmed by analytical calculations 
\citep[hereafter, TTW02]{ward97,ttw02} and numerical simulations \citep{npmk00,kdh01,dhk02,dkh03,blom03} in 
2D and 3D discs. It is well established that the driving force of planetary migration is exerted only at 
Lindblad and corotation resonances (GT80). While Lindblad torques excite and propagate density waves in 
discs \citep{ward97}, corotation torques do not. Due to the accumulation of angular momentum around the 
corotation region in discs, without any removal mechanisms such as disc viscosity \citep{ward91,m01,m02}, 
the corotation torque saturates (i.e. becomes ineffective).  

The problem of migration in the standard disc models is that, without some sort of slowing mechanism, 
this tidal interaction between a planet and the gaseous disc is so efficient that the migration timescale 
is much shorter than the disc lifetime ($\sim $ 1 - 3 Myrs). This fast mode of migration, known as 
type I migration, is applicable for low mass planets which cannot open up a gap in discs. Based on the core 
accretion scenario, which is the most accepted theory for planet formation, cores of gas giants and rocky 
planets are classified as type I migrators.  Massive planets such as gas giants can open a gap and undergo 
the so-called type II migration, which is about two orders of magnitude slower than type I migration 
\citep[e.g.][]{ward97}. Population synthesis models confirm that type I migration should be about two orders 
of magnitude slower than expected in order to qualitatively reproduce the observed mass-period relation 
\citep{il08v,mab09}. One of the greatest challenges in planetary formation and migration is to identify what 
physical process(es) succeeds in reproducing the diversity of exoplanets and under what conditions planetary 
systems similar to our Solar System are formed.

This long-standing problem can be resolved by paying close attention to the properties of discs. Tidal torque 
strongly depends on the distributions of gas and temperature in discs and is sensitive to inhomogeneities 
in these quantities \citep[hereafter, HP10, and references herein]{hp10}. Two sources of disc 
inhomogeneity have been proposed in the literature: dead zones and ice lines. Dead zones, regions with 
high density and low turbulence \citep[i.e. where the magneto-rotational instability (MRI) is quenched;][]{g96,mp06}, 
drastically affect these distributions and eventually 
provide two of the most robust slowing mechanisms \citep[HP10]{mpt07}. Both of them are the consequence of 
the difference in turbulence between the active and dead zones. One barrier arises by piling up gas at a 
dead zone's outer boundary which is a consequence of the time-dependent, viscous evolution of discs that have 
strong variations in the turbulent intensity \citep[$\alpha$;][]{mpt07,mpt09}. The other is the result of a 
positive temperature gradient (HP10). This radial temperature gradient in the dead zone is produced by the back 
heating from a dusty wall which is left in the active region due to the enhanced dust settling in the dead zone 
\citep[hereafter, Paper I]{hp09b}. The variation in turbulence with radial direction is the focus in the above 
treatments. The vertical variation in turbulence, resulting in layered structures: the MRI-active surface and 
the MRI-dead regions \citep{g96} is also important. Combined with the layered 
structures, the ice lines act as a barrier for type I migration \citep[e.g.][]{il08v}. This barrier 
arises because at the ice lines, the ice-coated dust density suddenly increases due to the low 
disc temperature. Consequently, the electron number density in the surface drops sharply due to the 
absorption by such dust. Since electrons are argents for gas to be MRI-active, the gas density of the 
MRI-active surface region falls and gas piles up in the MRI-dead region. As a result, a radial positive gradient 
for the surface density appears, which acts as a barrier for type I migration \citep{mmcf06,pp09a}. 

In this paper, we step back and investigate the general effects of dust settling on planetary migration in 
homogeneous discs. Although dust settling is observationally confirmed to be ubiquitous in discs around young 
stars (Paper I, references herein), there is no comprehensive study so far on the effect that this has 
on planetary migration. Furthermore, dust settling is one of the important processes for planet formation 
\citep[e.g.][]{bdb10}. In order to proceed, we compute detailed, radiatively heated disc models 
around an M star as described in Paper I, which include dust settling and the gravitational force of planets 
embedded in the disc. We have two main reasons to focus on M star systems. First, low-mass planets 
such as Super-Earths are the current preliminary targets for the ongoing and future observational missions such 
as Kepler. M star systems have the highest probability of detection using the transit method. In addition, some 
planets around M stars have already been found in or near the habitable zone. Second, Monte Carlo calculations 
become computationally expensive for more massive discs (such as those around classical T Tauri stars), making 
M star systems an ideal target.

We generalize our disc models by using minimum mass solar nebula (MMSN) 
models in which the surface density behaves as $\Sigma\propto r^{-3/2}$, as well as models with 
$\Sigma \propto r^{-1}$ \citep[hereafter S07]{sjw07} used in Paper I and HP10. We then compare the type I 
migration of various low mass planets (2, 5, 10 $M_{\oplus }$) within our models. Thus, our paper provides a 
detailed exploration of the effect of the thermal structure of discs, as a result of dust settling, on 
planetary migration. 

Our plan of this paper is the following. In $\S$ \ref{disk}, we describe and compute our disc models and summarize 
the important points for planetary migration for two disc models with different surface 
density profiles. We use these disc models as a background disc structure for analytical torque calculations. 
In $\S$ \ref{torque}, we briefly describe the Lindblad torque and discuss the corotation 
torque. In $\S$ \ref{results1} and $\S$ \ref{results2}, we present our results for the cases of well mixed 
and dust settling, respectively. We discuss the effects of individual and combined components such as dust 
settling and the gravitational force of planets. The effects of the surface density profiles on the 
migration are also discussed. In $\S$ \ref{discus}, we discuss other heat sources and corotation torques, 
which are both neglected in this paper. Finally, we discuss 
the effects of dust settling on type II migration. In $\S$ \ref{conc}, we present our conclusions. 

\section{Disc structures with dust settling} \label{disk}

We adopt the disc models developed in Paper I. In that paper, we computed the density and thermal structures of 
radiatively heated discs around an M star, by solving the radiative transfer equation by mean of a Monte 
Carlo method in 2D axisymmetric ($r$-$z$) discs. The main heat source is stellar irradiation that is 
absorbed mainly by dust in discs. In order to make our disc models realistic, the gravitational force of planets 
and dust settling are included. We describe the treatment of the gravity of a planet in the next subsection. 
In this paper, we adopt two disc models, one of which is an MMSN model ($\Sigma \propto r^{-3/2}$), 
and the other of which is an S07 model ($\Sigma\propto r^{-1}$). These two column density models bracket 
the likely range of models for observed protoplanetary discs. The primary mass is 0.1 $M_{\odot}$ and the 
disc mass is $4.5 \times 10^{-3} M_{\odot}$, which is increased by a factor of 10 compared with the typical 
disc mass around M stars in order to allow more massive planets to be formed \citep[e.g.][]{ambw05}. For the 
other disc parameters, we refer the reader to Paper I.

Note that our $r$-$z$ disc configurations are very useful for calculations of Lindblad torque which is well 
represented by analytical formulae while they are not for corotation which needs hydrodynamical simulations in 
$r$-$\phi$ discs. However, the corotation torque is at least 3 times smaller than the Lindblad torque in our 
disc models, which is discussed more in the next section. Thus, the inability of capturing corotation torque 
in our disc models is unlikely to change our findings significantly.

In this section, we briefly summarise the features of our disc models and the resultant temperature and 
density structures of discs. In addition, we discuss the differences between the two disc models, and elaborate 
on the grid systems we devised in order to accurately calculate Lindblad torque.

\subsection{The gravity of planets}

In this subsection, we briefly summarise the treatment of the gravitational force of a planet in our disc models. 
For complete discussion, we refer the reader to Paper I. We incorporate the gravitational force of a planet into 
the standard vertical hydrostatic equilibrium equation, which may be written as
\begin{equation}
\rho=\rho_0 \exp \left( -\frac{z^2}{2h^2} + \mu \frac{r^2 }{h^2} \left[\frac{r}{\sqrt{(r-r_p)^2+z^2}}-
\frac{r}{|r-r_p|} \right]  \right),
\label{rho_w_pl}
\end{equation}
where $\mu=M_p/M_*$, $M_p$ is the planetary mass, $M_*$ is the stellar mass, $r_p$ is the location of the 
planet from the host star, $h$ is the disc scale height, and the normalization constant $\rho_0$ is chosen 
so that the density at $z=0$ corresponds to the unperturbed density (following \citet{js03}). Thus, the dust and 
gas distributions are perturbed by the presence of a planet, and consequently the thermal and density structures 
are affected by the planet.

We interpolate the density and temperature of discs in the region within the Hill radius 
$r_H\approx r_p(M_p/M_*)^{1/3}$ following \citet{js05} because the hydrostatic assumption breaks down within 
the region. It is well known that the horseshoe structure and/or circumplanetary discs are located well inside 
the Hill radius \citep[e.g.][]{cbkm09}, so that we consider this treatment to be very conservative since 
use of the Hill radius overestimates the effect.
This interpolation is unlikely to be harmful for calculating the Lindblad torque because the resonant positions 
are pushed away from their distance from the planets $2h/3\approx r_H$. 

\subsection{Grid systems}

The Lindblad torque takes its maximum value at the distance $2h/3$ from a planet due to the gas pressure 
effect. It then decreases with increasing distance from the planet \citep{ward97}. Thus, it is important to 
highly resolve the thermal structure of the region in the vicinity of planets in order to calculate 
the Lindblad torque accurately. The thermal structure of discs, however, is determined globally since 
discs are heated by their central stars. Furthermore, the higher the resolution, the longer is the 
computational time. This is because a larger number of photons is required in finer grid systems to avoid 
random noise produced by the Monte Carlo methods. Therefore, carefully constructed grid systems are required 
to optimize the computational time.

We adopt a logarithmic grid system in vertical direction. This is sufficient because the inner region in 
which the torques provide the largest contribution is efficiently resolved by finer grids. In the radial 
direction, however, more complicated grid systems are required. We combine two types of grid 
systems. Around the planet, we use an equally spaced, fine mesh system while for the rest of the disc, we use a 
logarithmically spaced, coarser mesh. The fine mesh is inserted into the vicinity of the planets with its size 
$\bigtriangleup R$ centered at the location of the planet, while the larger remaining region is resolved by 
the coarser mesh. We fix the disc size as 50 au and the total number of grids as 180. Also, the region 
$\bigtriangleup R$ is resolved by 65 grids for any case.

Furthermore, we recalculate the density structure after temperature calculations are completed by using 
$n$ times finer systems mentioned above. This treatment enables one to reduce the computational time by a 
factor of 10 or so. We performed a convergence study for the Lindblad torque in our grid system by changing 
$\bigtriangleup R$  and $n$, and found $\bigtriangleup R=6h$ and $n=6$ are sufficient \citep{h08}.
  
\subsection{Well mixed dust}

\begin{figure*}
\begin{minipage}{17cm}
\begin{center}
\includegraphics[height=4.2cm]{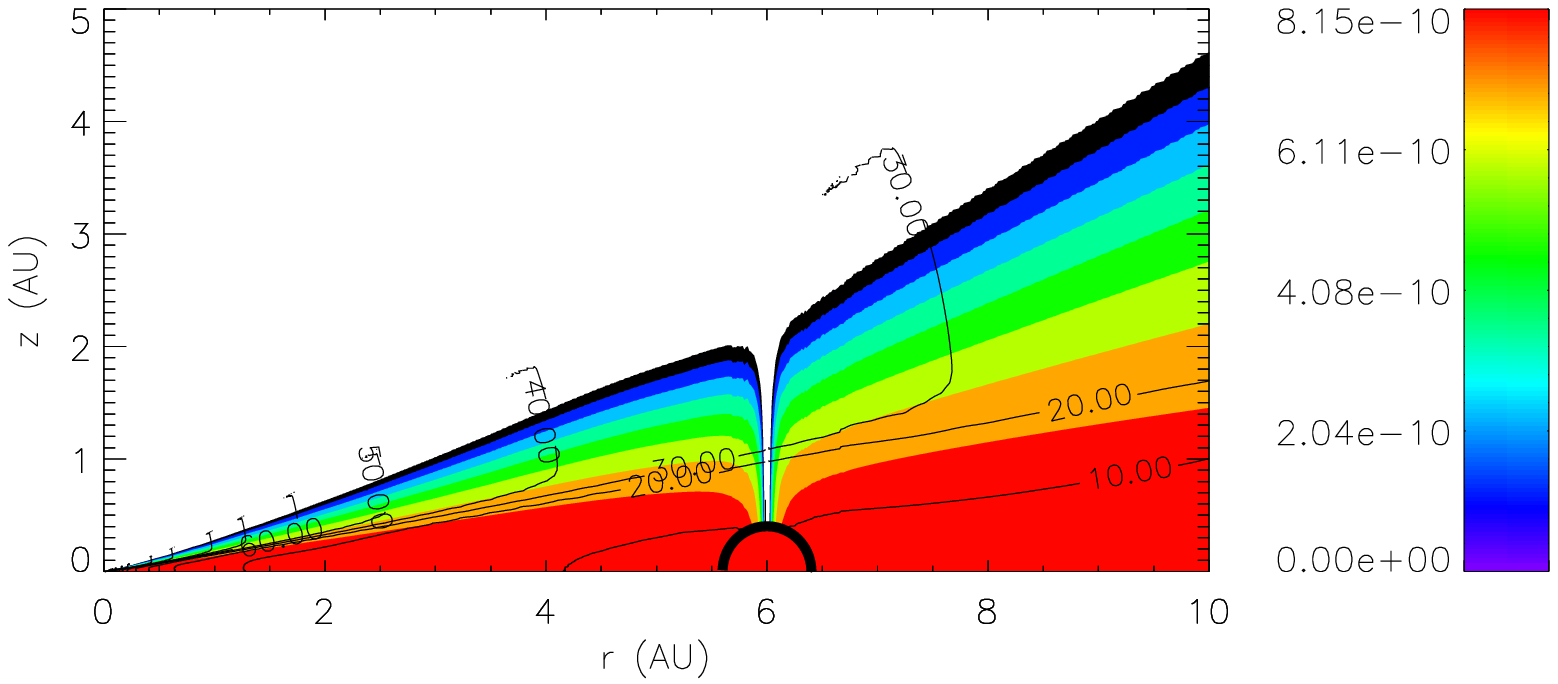}
\includegraphics[height=4.2cm]{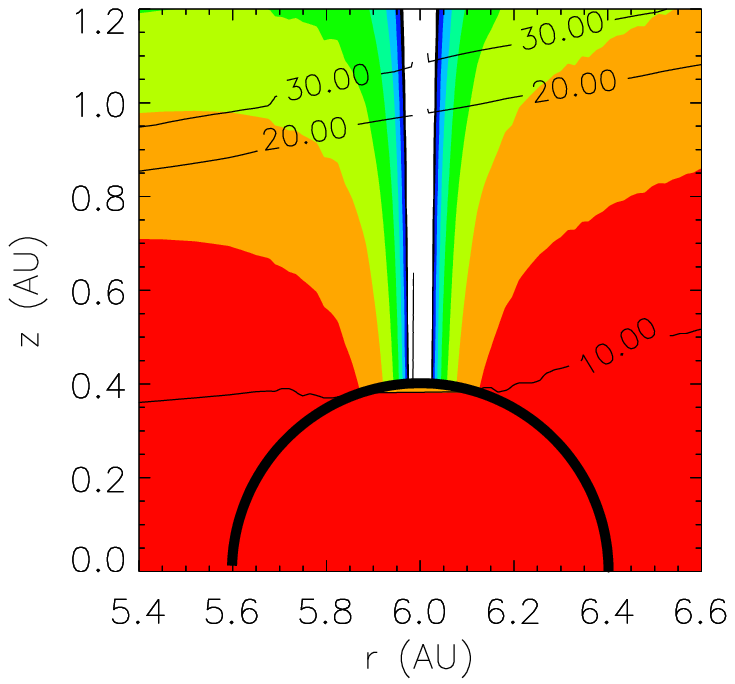}
\includegraphics[height=4.2cm]{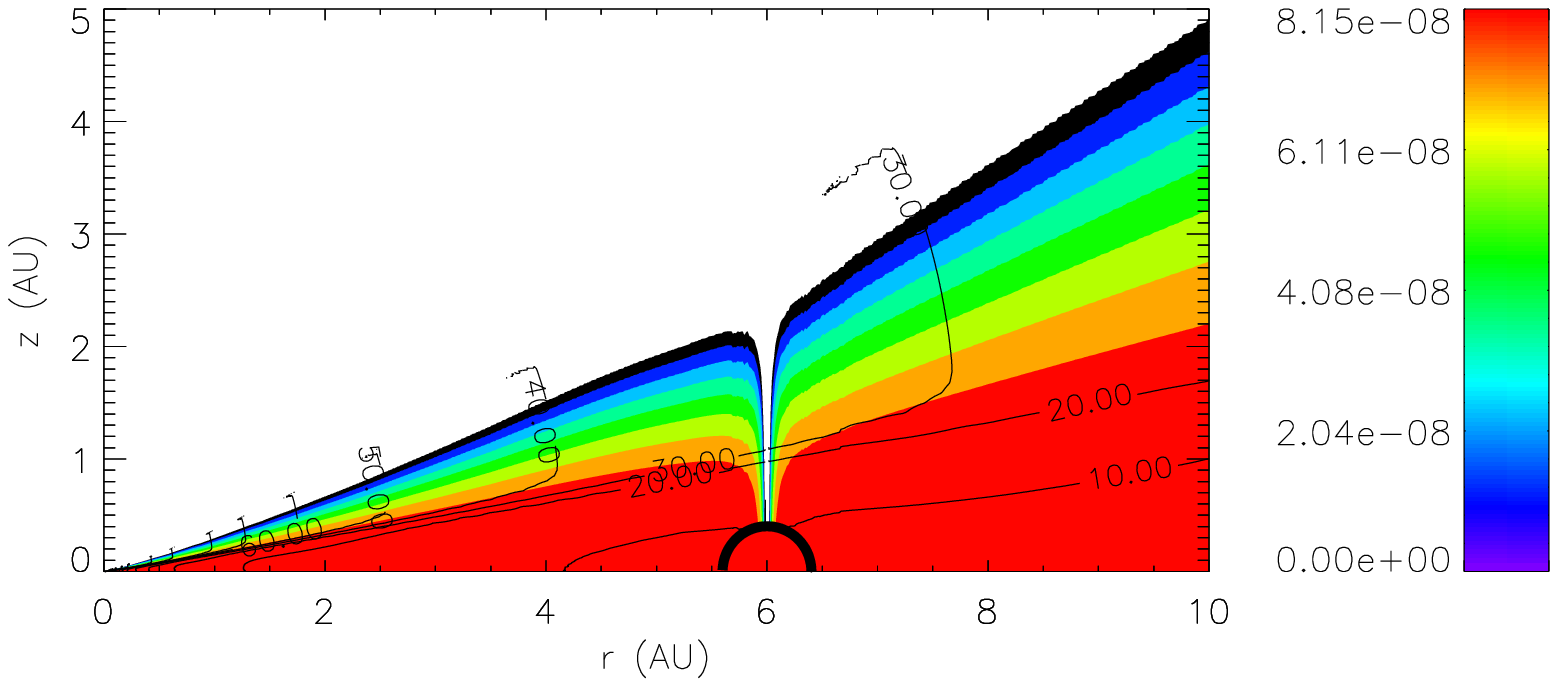}
\includegraphics[height=4.2cm]{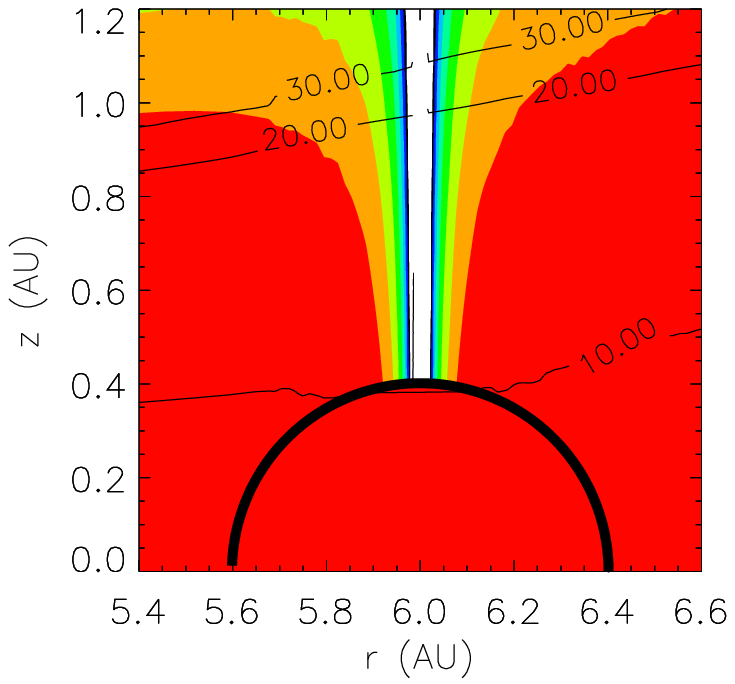}
\includegraphics[height=4.2cm]{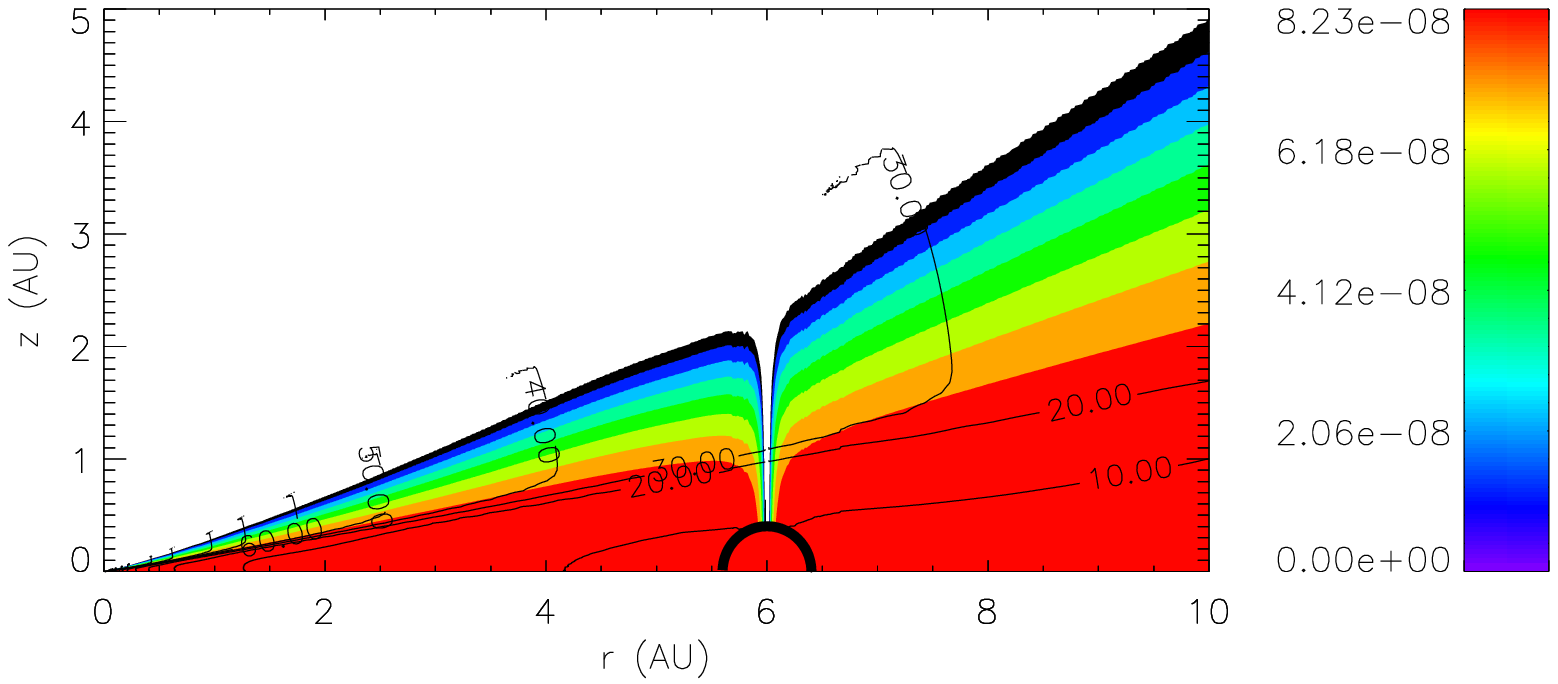}
\includegraphics[height=4.2cm]{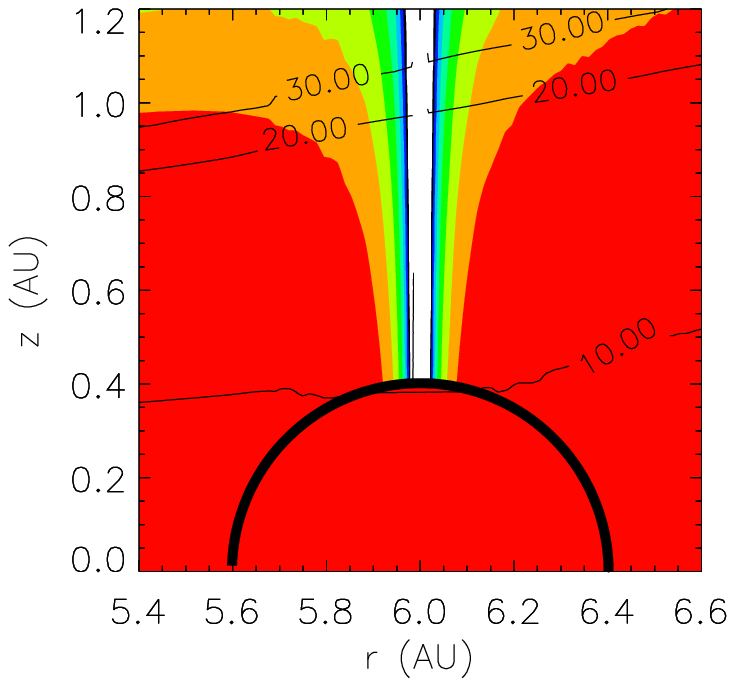}
\caption{Fully mixed dust models. The density and temperature structures of a MMSN disc model with a 10 
$M_{\oplus }$ mass planet at 6 au. The density is denoted by the colors. The scale is shown in the color 
bar in units of [g cm$^{-3}$]. The temperature is denoted by the contours in the unit of [K]. 
Top: dust density. Middle: gas density. Bottom: total density (gas + dust). Blow-ups of the region 
near the planet are shown in the right column. The thick solid lines denote the Hill radius of 
the planet on every panel. Note that the disc temperature shown on every panel is identical to each 
other because it is calculated as the mass-averaged dust temperatures. The surface layer has higher 
temperatures while the mid-plane region has lower temperatures. The density above the planet is reduced 
by its gravitational force. This results in higher temperatures and low densities above it.}
\label{fig1}
\end{center}
\end{minipage}
\end{figure*}

Dust is important for heating the disc. We first 
summarise the results of our simulations of the thermal and density structures of discs with a planet for 
the case of well mixed dust. Fig. \ref{fig1} shows the dust, gas and total (gas + dust) density 
distributions of the MMSN disc models respectively on the top to bottom panels. A 10 $M_{\oplus }$ planet 
is placed, without loss of generality, at 6 au in Fig. \ref{fig1}. Also, the zoomed-in versions of disc 
structures around the planet are shown in the right column. Note that the disc temperatures shown on every 
panel are the same because they are defined as the mass-averaged temperature of dust with various grain 
sizes (Paper I). 

\begin{figure}
\begin{center}
\includegraphics[width=8.4cm]{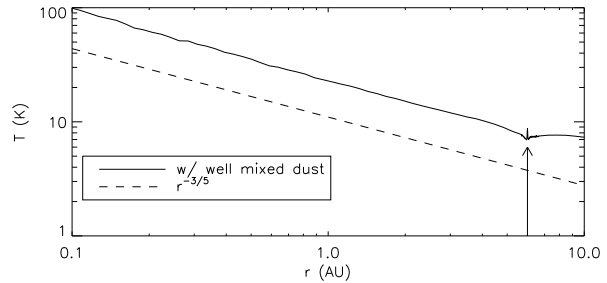}
\caption{Well mixed dust models. The temperature structure as a function of disc radius for a MMSN disc model with 
a 10 $M_{\oplus }$ mass planet at 6 au. The solid line denotes the temperature of our calculations. 
The dashed line denotes the best fit to them. A sharp peak is produced at the location of the planet 
by the compression arising from the gravitational force of the planet, indicated by an arrow. Also, 
the compression results in lower temperatures around it.}
\label{fig2}
\end{center}
\end{figure}

Our numerical simulations confirm that the mid-plane region has much lower temperatures than the surface 
layer for both disc models (Fig. \ref{fig1}). We found, for both disc models, that the temperature of the 
surface is $\simeq 80$ K at 1 au while that of the mid-plane is $\simeq 25$ K there. This arises because 
the surface layer is directly heated by the central star while the mid-plane region is heated only by the 
thermal emission of dust \citep{cg97}. Also, the direct heating of discs results in geometrically flared 
structures, which are seen in both the dust and gas density distributions. Furthermore, the presence of the 
planet produces a low density region above it due to the compression by the gravitational force of the 
planet. Consequently, the region above a planet has higher temperatures while the mid-plane region, in turn, 
has lower temperatures. The compression results in a small localized peak in a temperature profile in the 
mid-plane (Fig. \ref{fig2}).

The difference between the MMSN and S07 disc models is that the temperature in the mid-plane region becomes 
lower in the former model, especially at the region around the planet and larger disc radii. We 
found that the temperature difference between two disc models is about 15 per cent in the mid-plane region 
at 10 au. This is explained by the two combined effects arising from a steeper slope profile. The MMSN disc 
model has a denser inner region, and is therefore more optically thick. In our disc setup, 
the MMSN model is about 4 times denser than the S07 model in the mid-plane region at 1 au. As a result, the 
inner region in the MMSN disc model prevents photons from readily penetrating the outer region, resulting 
in lower temperatures there. The higher density in the inner region, in turn, results in a less dense, outer 
region even in the mid-plane region, since the total disc mass is fixed. The outer region consequently 
becomes optically thin even to the thermal emission of dust, resulting in lower temperatures. Since the 
difference of the optical depth is accumulated with increasing disc radius, the temperature difference 
far from the star becomes larger. It is interesting that both disc models result in the almost identical 
temperature profiles at the mid-plane, especially at smaller disc radii, and are well represented by 
$r^{-3/5}$ (Fig. \ref{fig2}).

\subsection{Dust settling} \label{disk_ds}

\begin{figure*}
\begin{minipage}{17cm}
\begin{center}
\includegraphics[height=4.2cm]{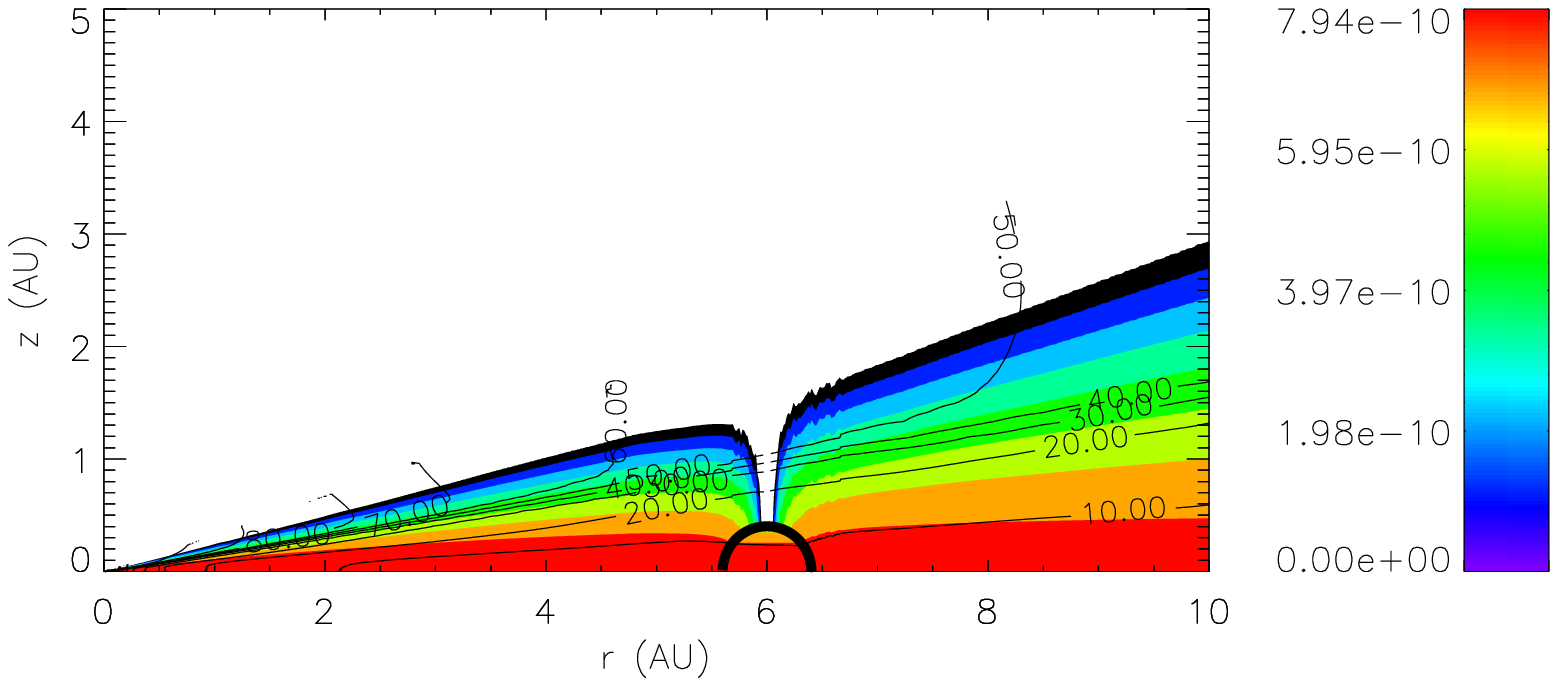}
\includegraphics[height=4.2cm]{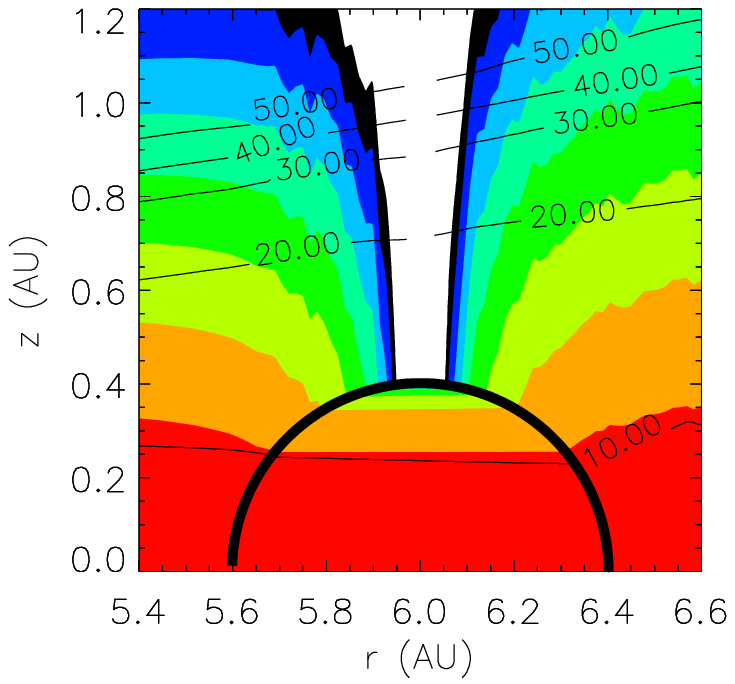}
\includegraphics[height=4.2cm]{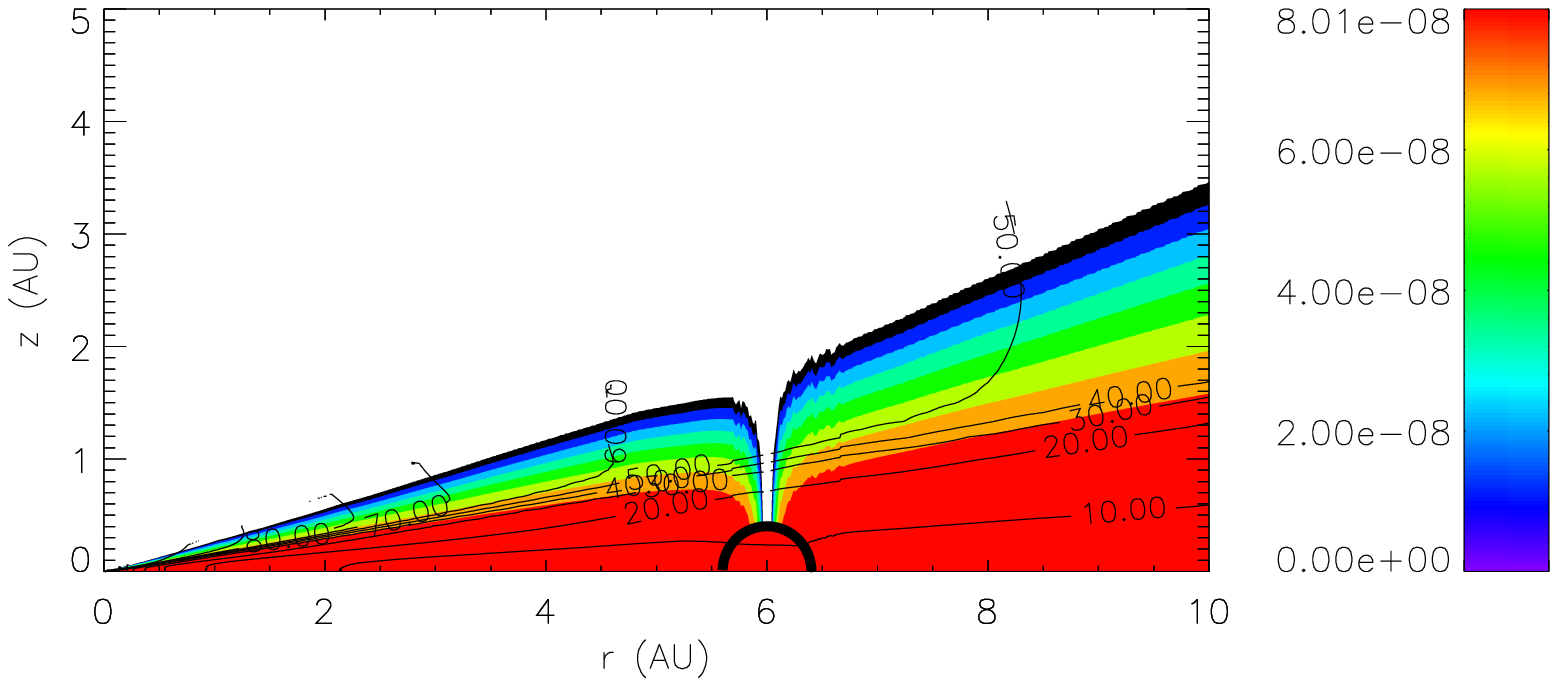}
\includegraphics[height=4.2cm]{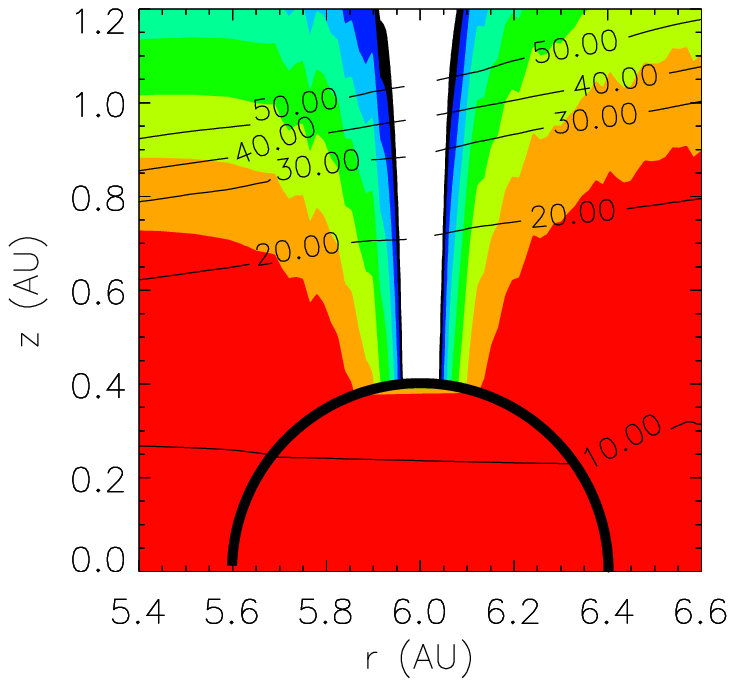}
\includegraphics[height=4.2cm]{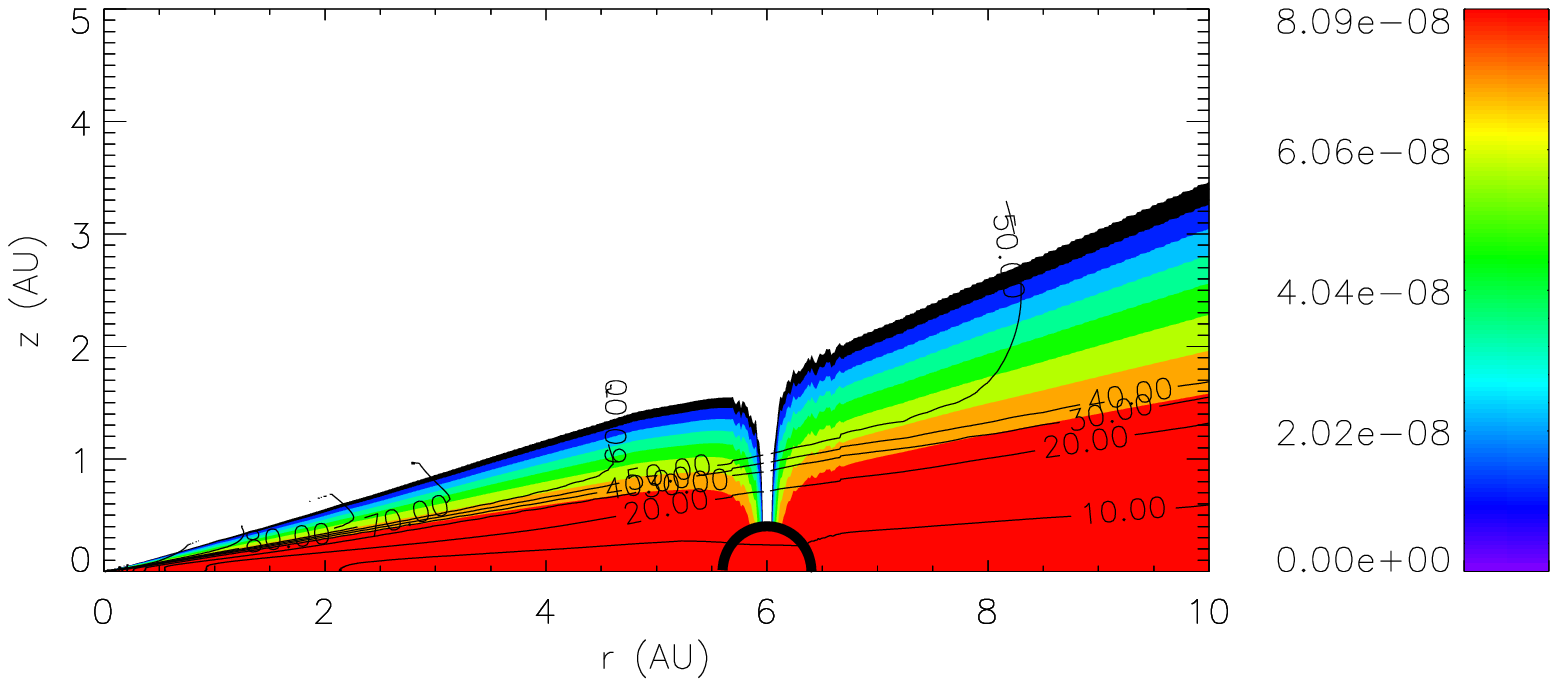}
\includegraphics[height=4.2cm]{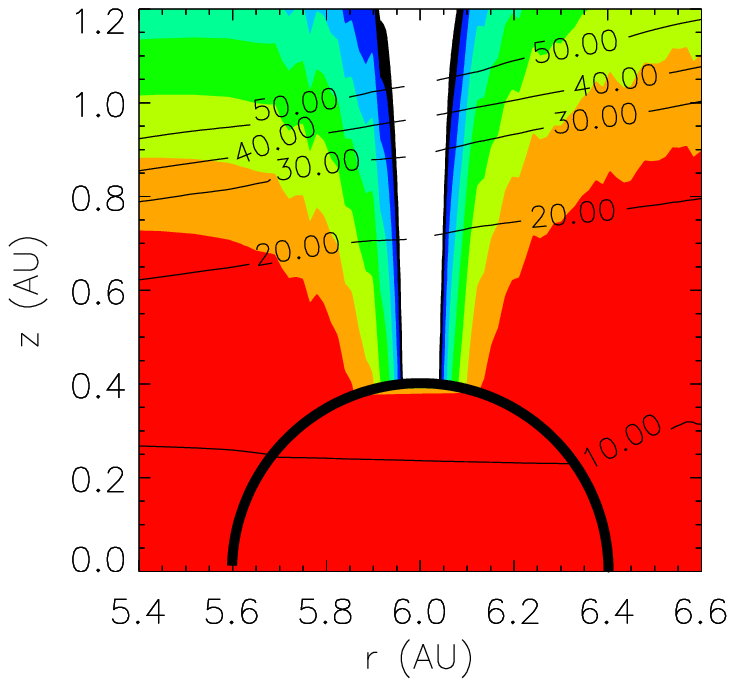}
\caption{Dust settling models. The density and temperature structures of a MMSN disc model with a 
10 $M_{\oplus }$ mass planet at 6 au (as Fig. \ref{fig1}). Top: dust density. Middle: gas density. 
Bottom: total density (gas + dust). Dust settling makes the surface layer hotter and the mid-plane region 
cooler, resulting in a geometrically flatter disc shape. The presence of the planet provides similar 
effects on the density and temperature structures of the disc.}
\label{fig3}
\end{center}
\end{minipage}
\end{figure*}

\begin{figure}
\begin{center}
\includegraphics[width=8.4cm]{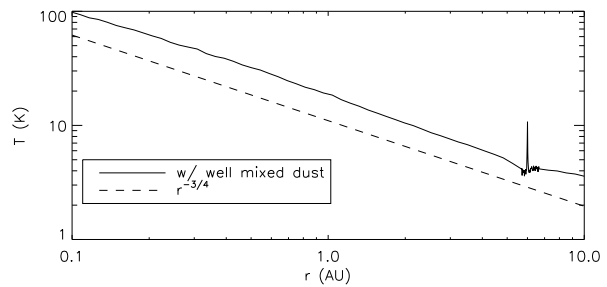}
\caption{Dust settling models. The temperature structure as a function of disc radius for a MMSN disc model with 
a 10 $M_{\oplus }$ mass planet at 6 au (as Fig. \ref{fig2}). The best fit profile is identical to an analytical 
derivation for flat discs. A sharp peak at the location of the planet is enhanced due to the combined 
effects of dust settling and the presence of the planet.}
\label{fig4}
\end{center}
\end{figure}

We showed that dust settling drives the disc structure from flared to flatter shapes in Paper I. Fig. 
\ref{fig3} shows that the same behaviour pertains to the MMSN model as well. Dust settling arises from the 
size distribution of dust grains. In general, larger grains cannot be kept aloft even within a region with 
a high level of turbulence ($\alpha=10^{-2}$) because collisions with gas become inefficient. The increase 
of larger grains in the mid-plane region strongly reduces the probabilities for smaller grains there to be 
exposed to and to absorb photons emitted by the star. In the surface layer, however, the probability for 
small grains to interact with the photons does not change very much since this region is originally optically 
thin. Taking the mass-average of the dust temperatures for all sizes, the surface layer has higher 
temperatures and the mid-plane region has lower temperatures, compared with the well mixed case. These 
lower temperatures in the mid-plane, in turn, reduce the scale height of gas, resulting in a flatter 
shape of gas. Table \ref{table1} shows the variations of the scale height, the temperatures of the surface and 
mid-plane due to dust settling in per cent at 1 au for two disc models. Furthermore, geometrically flatter 
disc shapes are favoured because of the temperature of the mid-plane, which is well-represented 
by $r^{-3/4}$ (Fig. \ref{fig4}). This profile is identical to analytical derivations for flat discs 
\citep{cg97}. 

Dust settling does not change the effects of the gravitational force of the planet for the two disc models. 
Thus, the gravity of the planet produces higher temperatures above it and lower temperatures in the mid-plane 
region. A sharp peak at the co-orbital radii of the planet is enhanced by the combined effects of dust 
settling and the planet for both disc models (Fig. \ref{fig4}).

The difference arising from the two power law behaviours of the surface density ($s=-1$ vs -3/2) is 
qualitatively similar to that found for the well-mixed dust case (see table \ref{table1}). Since dust 
settling makes discs flatter and increases the optical depth of the mid-plane region considerably, 
the difference arising from different slopes of the surface density is diminished, compared with the well 
mixed case. We found that the temperature difference between two disc models reduces from 15 to 6 per cent 
in the mid-plane region at 1 au, when dust settling is taken into account. Thus, dust settling reduces the 
interaction with photons emitted from the star in the mid-plane region for both disc models while it keeps 
the optical depth thick enough for the thermal emission of dust even for the MMSN models. 

\begin{table*}
\begin{minipage}{17cm}
\begin{center}
\caption{The variation of the disc properties due to the dust settling at 1 au}
\label{table1}
\begin{tabular}{ccc}
\hline
                                        & MMSN disc models & S07 disc models    \\ \hline
the scale height $h$                    & 10 \% reduced    & 15 \% reduced      \\
The temperature of the surface region   & 28 \% increased  & 28 \% increased    \\
The temperature of the mid-plane region & 20 \% reduced    & 28 \% reduced      \\
\hline
\end{tabular} 
\end{center}
\end{minipage}
\end{table*}

\section{Tidal torque} \label{torque}

We focus on type I migration because it endangers the survival of rocky planets and cores of gas giants in 
the standard disc models. We discuss type II migration in $\S$ \ref{typeII}. We assume the mass of planets 
to be small enough to perform type I migration, but to be large enough to disturb the surrounding disc. 
We neglect any other back-reaction from planets onto the disc structure. Thus, we can reasonably estimate 
the migration time analytically for any given disc structure. In this paper, we assume planets to be in 
circular motion in order to simply estimate the migration time. In addition, we exclude corotation torque 
because it is safely neglected in our disc models as discussed below.

\subsection{Torque at the Lindblad resonances}

Since the basic formula of the Lindblad torque is well discussed in the literature, we summarise our 
assumptions and modifications. The complete description is presented in Appendix \ref{app1}. We adopt 
the analytical formula of the Lindblad torque derived by \citet[hereafter, JS05]{js05}. In the formula, the 
(total) Lindblad torque $\Gamma^L(r_p)$ is assumed to be calculated by 
\begin{equation}
\Gamma^L(r_p)=\int^{\infty}_{-\infty}dz \frac{d \Gamma^L(r_p,z)}{dz},
\label{total_torque}
\end{equation}
where $d \Gamma^L(r_p,z) /dz$ is the Lindblad torque exerted by a layer at the height from the mid-plane $z$. Since 
the mid-plane region has the largest gas density, the torque arising there provides the largest contribution 
to the total. Furthermore, dust settling transforms the vertical structure of discs from flared to flatter 
shapes, and consequently it strengthens the torque by accumulating the gas density in the mid-plane region 
(see $\S$ \ref{results2}). We note that only the gas density is used for the torque calculations. One might 
suppose that the torque arising from the dust density is comparable to that from the gas for the dust settling case, 
but this is not the case, because the dust density is about 2 orders of magnitude smaller than that of gas 
(see Fig. \ref{fig3}). 

The above treatment is very useful for mimicking the reduction of the tidal torque in 3D discs, which is found 
by TTW02. In the other words, the integration effectively accounts for the dilution of the gravitational force of 
planets for the vertical direction without needing to rigorously solve the 3D Euler equations \citep{ward88,a93}. 
Similar approaches have been adopted by \citet[hereafter, MG04]{mg04} in which the gravitational force of planets is 
weakened by the disc scale height $h$. Thus, this formula includes only 2D modes and relies on the 
assumption that the vertical modes arising from a purely 3D effect are negligible. This assumption is 
also supported because \citet{lo98} who investigated the excitation and propagation of density waves at the Lindblad 
resonances in thermally stratified discs, and showed that 2D modes carry out more than 95 per cent of angular 
momentum and vertical modes are not dominant. 

For practical purpose, we modified the JS05 torque formula. In this modification, the torque density (not 
the torque itself) is calculated, so that one can avoid calculating the resonant positions \citep{ward97}.
Consequently, $d \Gamma^L(r_p,z) /dz$ is given as
\begin{equation}
\frac{d \Gamma^L(r_p,z)}{dz} =\int^{\infty}_{0}dr \frac{d}{dz} \left( \frac{d\Gamma^L}{dr}(r,z) \right),
\label{layer_torque}
\end{equation}
where $d^2 \Gamma^L(r,z)/dzdr$ is the torque density at $r=r$ and $z=z$. Under this modification, the 
Lindblad resonant positions $r$ are incorporated into the wave number $m$ (see equation (\ref{wave_num})). 
Since the resonant positions or $m$ are controlled by the gas pressure \citep{a93}, the lower temperatures 
of the mid-plane region induced by dust settling push the resonant positions closer to the planet. 
Consequently, one expects the Lindblad torque becomes stronger, which we will confirm 
(see $\S$ \ref{results2}).

\subsection{Torque at the corotation resonances}

The torque arising from the corotation resonances is trickier to compute because of the possibility of 
saturation (i.e. it becomes zero) as well as having an interaction region that is characterized by horseshoe 
orbits. However, our preliminary calculations showed that the unsaturated corotation torque in our disc 
models is at least 3 times smaller than the Lindblad torque \citep{pbck09}. In this calculations, we adopted 
the torque formula consisting of the linear Lindblad torque and both the entropy- and vortensity-related 
horseshoe drags (see their equation (47)). We assumed the surface density $\Sigma \propto r^{-3/2}$ or 
$\Sigma \propto r^{-1}$ and the disc temperature $T \propto r^{-1/2}$ with $\gamma=1.4$. With the reduction 
of the corotation torque due to (partial) saturation, it is reasonable to consider that migration is controlled 
only by the Lindblad torque (see Appendix \ref{app2} for complete discussion). For these reasons, we assume 
that the corotation torque can be safely neglected to zeroth order in our calculations. 

\subsection{Migration rate \& time}

We calculate the migration rate and time analytically by using the tidal torques above, following MG04 and 
JS05. Thus, our calculated migration rate and time are instantaneous values rather than time-dependent 
ones. Since we focus on type I migration and assume that the back-reaction from a planet on the disc 
due to the total tidal torque is negligible, the rate of change of angular momentum of the planets is
\begin{equation}
\frac{dJ}{dt}=-\Gamma^L,
\end{equation}
where $J=M_p\sqrt{GM_* r_p}$ is the angular momentum of the planet. Note that a minus sign appears since 
the planets lose angular momentum by exerting the tidal torque $\Gamma^L$ on the discs.

Since the radial drift is due to the migration, we assume $r_p=r_p(t)$. The migration rate is written as 
\begin{equation}
\frac{1}{r_p}\frac{d r_p}{dt}=-\frac{2\Gamma^L}{J}.
\end{equation}
Since $\Gamma^L$ is generally positive, $dr_p/dt$ becomes negative, implying that planets migrate inward. 
When $\Gamma^L$ becomes negative, the migration is outward.

The migration time $T_{mig}$ is estimated by
\begin{equation}
T_{mig} \approx  J \left(- \frac{dJ}{dt} \right)^{-1} = \frac{M_p\sqrt{GM_* r_p}}{\Gamma^L}.
\label{t_mig}
\end{equation} 
Again, the migration time becomes positive with $\Gamma^L$ positive, resulting in inward migration. 

\section{Results for the well mixed dust case} \label{results1}

We present our results for the well mixed case for our two disc models in this section. The density and 
thermal structures of discs discussed in $\S$ \ref{disk} are used as background disc structures for analytical 
calculations. 

\subsection{Torque}

In this subsection, we focus on the behaviour of the torque, $ d \Gamma^L (r_p,z)/ dz$, exerted by a layer of 
gas at height $z$ (see equation (\ref{layer_torque})), in order to examine the effects of planetary gravity on it. 
Since we found that the basic features of the results for the two disc models are similar, we first discuss the 
results for the case of S07 ($\Sigma \propto r^{-1}$) in detail and then compare the difference between them. 

Fig. \ref{fig5} shows the torque $d \Gamma^L/ dz$ as a function of $z$ (distance from the mid-plane) for 
the S07 disc models. Except for a small dip, the torque $d \Gamma^L/ dz$ of all planets monotonically 
increases toward the mid-plane. This trend arises partly because the temperature decreases toward the 
mid-plane and partly because the gravitational force of a planet is diluted for the vertical direction. 
The former reduces the gas pressure with decreasing $z$, causing the resonances to move closer to the planet. 
The latter increases the magnitude of the torque with decreasing $z$. Thus, both physical behaviours make 
the torque of a layer stronger with decreasing $z$. The presence of the dip is explained by the distortion of 
disc structure by the presence of the planet, which is discussed below.

As expected, the most massive planet exerts the largest tidal torque on the disc (but its mass is still 
less than the threshold mass to open a gap). Consequently, the planet performs the fastest migration 
(Fig. \ref{fig7}). As planetary masses decrease, the tidal torque becomes weaker, resulting in slower 
migration.

When the MMSN disc model is adopted, the position of the dip in the torque moves slightly upward. We found 
that the torque $d \Gamma^L(r_p,z) /dz$ for the MMSN disc model is a steeper function of $z$. In the other 
words, it becomes stronger in the mid-plane region while it becomes weaker in the surface layer 
(also see the left column of Table \ref{table2}). This occurs because the MMSN model results in lower temperature 
structures around the mid-plane region. In addition, these lower temperatures cause the gas distribution in the 
disc to be slightly less flared since the gas distribution is determined by the hydrostatic equilibrium (which is 
controlled by the disc temperature in the mid-plane). 

In order to better understand the presence of the dips, we compute the 
scaled torque exerted by a layer of gas at height $z$, following JS05. JS05 first studied the torque behaviour 
in radiatively heated discs in which the influence of planetary gravity is taken into account. The scaled torque 
is defined as 
\begin{equation}
\frac{d \Gamma^L_{scal}}{dz}=\frac{1}{\Gamma_0}\int_0^{\infty} dr 
                                    \frac{d}{dz} \left( \frac{\Sigma}{\rho} \frac{d\Gamma^L}{dr} \right),
\label{scaled_torque}
\end{equation}
where 
\begin{equation}
\Gamma_0=\left( \frac{M_p}{M_*} \right)^2 \Sigma_p r_p^4 \Omega_p^2,
\end{equation}
and where any quantity with suffix $p$ takes a value at $r=r_p$. The scaled torque is designed to isolate the 
temperature dependence of the torque from each layer, since the effects of the density distortion on the torque are 
canceled out by the term $\Sigma/\rho$ (see equation (\ref{scaled_torque})). 

Fig. \ref{fig6} shows the scaled torque as a function of $z$ (distance from the mid-plane) for the S07 disc models. 
As with the case of the torque $d \Gamma^L / dz$, the scaled torque of all planets monotonically increases toward 
the mid-plane, except for the dip. This is consistent with JS05. Due to the $\Gamma_0$ in equation 
(\ref{scaled_torque}), the scaled torques for all planets are normalised to the same value at $z=0$. It is 
interesting that the structure of the dips 
is enhanced in the scaled torque. This enhancement provides a clear demonstration that dips are 
located slightly below $z=r_H$ (which is denoted by the vertical lines). This is similar to the findings of JS05. 
We can explain the dip formation, as follows. The compression of the material above 
the planet due to its gravity, results in higher temperatures in the region by reducing the optical depth there. 
This resultant temperature structure pushes the resonant positions further away from the planet and eventually 
decreases the torque there. Thus, we can conclude that the presence of the dip is explained by the distortion of 
disc structure by the presence of the planet. 

Since the temperature distortion which arises from the density distortion becomes larger with the mass of the 
planet, more massive planets exert a weaker scaled torque on the disc. Thus, the temperature effects arising from 
the distortion diminish the (scaled) torque which is reduced even more by massive planets.

We emphasize that the compression of the material above planets, in turn, lowers the temperature of the mid-plane 
region (see $\S$ \ref{disk}). This makes the torque stronger there. Fig. \ref{fig6} shows that the scaled torques 
of planets with different masses are almost identical around the mid-plane region. This implies that the 
temperature variation around the mid-plane is very insensitive to planetary mass. Since the total torque is 
calculated by summing $d \Gamma^L /dz$ from each layer (see equation \ref{total_torque}), the effects of planets 
on the torque are determined by the decrement created by the dips as well as the increment around the mid-plane. 
In our both disc models, the increment dominates over the decrement.  

As with the results of the torque $d \Gamma^L/ dz$, the main differences between two disc models (MMSN vs S07) are 
the position of the dips and the steepness of the scaled torque as a function of $z$. These differences arises 
from the change in the temperature structure discussed in $\S$ 3.2, that is, the steeper surface density slope 
provides lower temperatures around the planet and the mid-plane region. Except for these, the basic features are 
the same as above. 

\subsection{Migration time \& rate}

In both disc models, massive planets migrate more quickly than low mass planets. This trend is clearly shown in 
Fig. \ref{fig7} which shows the migration time and rate as a function of orbital radius for planets with various 
masses (top and bottom panels, respectively). The migration time increases for planets with large orbital radii, 
as expected. Also, migration in the MMSN disc models is more rapid relative to the S07 disc models, as discussed 
above. We found that migration in the MMSN disc models is up to 4.0 times faster at 1 au than in the S07 disc 
models (see the left column of Table \ref{table2}). 

It is interesting that the migration time for planets in discs around M stars is comparable to that of 
classical T Tauri systems. This can be shown by making an order of magnitude estimate. The tidal torque is scaled by 
$\Sigma_p (r_p/h_p)^2$ \citep[TTW02;][]{pbck09}.  At 1 au, discs around classical T Tauri systems typically have 
$\Sigma_p (r_p/h_p)^2 \simeq 10^4$ g cm$^{-2} \times (1/0.1)^2 = 10^6$ \citep{cg97}. For discs around M stars, 
$\Sigma_p (r_p/h_p)^2 \simeq 10^2$ g cm$^{-2} \times (1/0.01)^2 = 10^6$ (S07). Thus, the planetary migration 
time for discs around both stars is similar. Note that, in these above simple calculations, we used 
characteristic values of $r_p/h_p$ to estimate an order of the timescale.  

\begin{table}
\caption{The difference in the migration time between the S07 and MMSN disc models at 1 au}
\label{table2}
\begin{tabular}{ccc}
\hline
                 & well mixed & dust settling \\ \hline
2 $M_{\oplus}$   & 4.0        & 1.4 (1.5)     \\
5 $M_{\oplus}$   & 4.0        & 1.4 (1.5)     \\
10$M_{\oplus}$   & 3.4        & 1.5 (1.8)     \\
\hline
\end{tabular} 

The maximum difference found in discs is given in the brackets if the number in the table is not
\end{table}

\subsection{Mass dependence}

In order to further elucidate the effects of the mass of the planet, we replot the migration time as a 
function of planetary mass at two different orbital radii (Fig. \ref{fig8}). If these effects arising from 
the gravitational force of a planet were neglected, the migration time would scale as an inverse function of their 
masses denoted by the dotted lines (also see equations (\ref{t_mig}), (\ref{torque_density}) and 
(\ref{torque_tot})). Obviously, our calculations show some deviation from this function for both disc models 
(although it is not so large). The deviation increases with planetary masses, resulting in faster migration. 
This indicates that the increment of the torque around the mid-plane is larger than the decrement by the dip 
formation around $z=r_H$, and that the difference increases with planetary mass. 

JS05 found that the deviation increases with planetary mass, but it results in slower 
migration. This indicates that, in their disc models, the decrement is larger than the increment. It is well known 
that the dip formation strongly depends on the grazing angle at which photons emitted from the star strike the disc 
surface \citep{js04}. For the discs around the classical T Tauri stars JS05 examined, the scale height is large, and 
consequently the grazing angle is also large, compared with the discs around M stars we examined. As a result, the 
decrement by the dip formation is enhanced. 

In addition, the deviation becomes smaller with increasing orbital radius of the planet. This is 
understood by the power-law function of the surface density. Since the disc density decreases with the 
distance from the central star, the density and temperature distortions caused by the gravitational force 
of the planet are diminished at larger disc radii. Consequently, the migration time approaches the simple 
analytical solution, which is different from JS05. We suggest that the main reason for the difference 
is that they targeted massive discs around classical T Tauri stars including viscous heating 
(although our target is low mass discs around an M stars without viscous heating). In addition, the orbital 
radii of the planet they considered are 0.5 to 4 au. For such a short range, the effects arising from the 
planet on the disc are not diminished. Our results, by contrast, pertain to geometrically thin, low mass discs and 
the outer part of massive discs for which viscous heating is safely neglected while the results of JS05 are 
for geometrically thick, (the inner part of) massive discs in which viscous heating is not negligible.

It is interesting that the deviation from $1/M_p$ is almost identical for both disc models if the migration 
times are appropriately scaled. Also, the difference of the migration times between two disc models becomes 
small with increasing the orbital radius of the planets.
 
\begin{figure}
\begin{center}
\includegraphics[width=8.4cm]{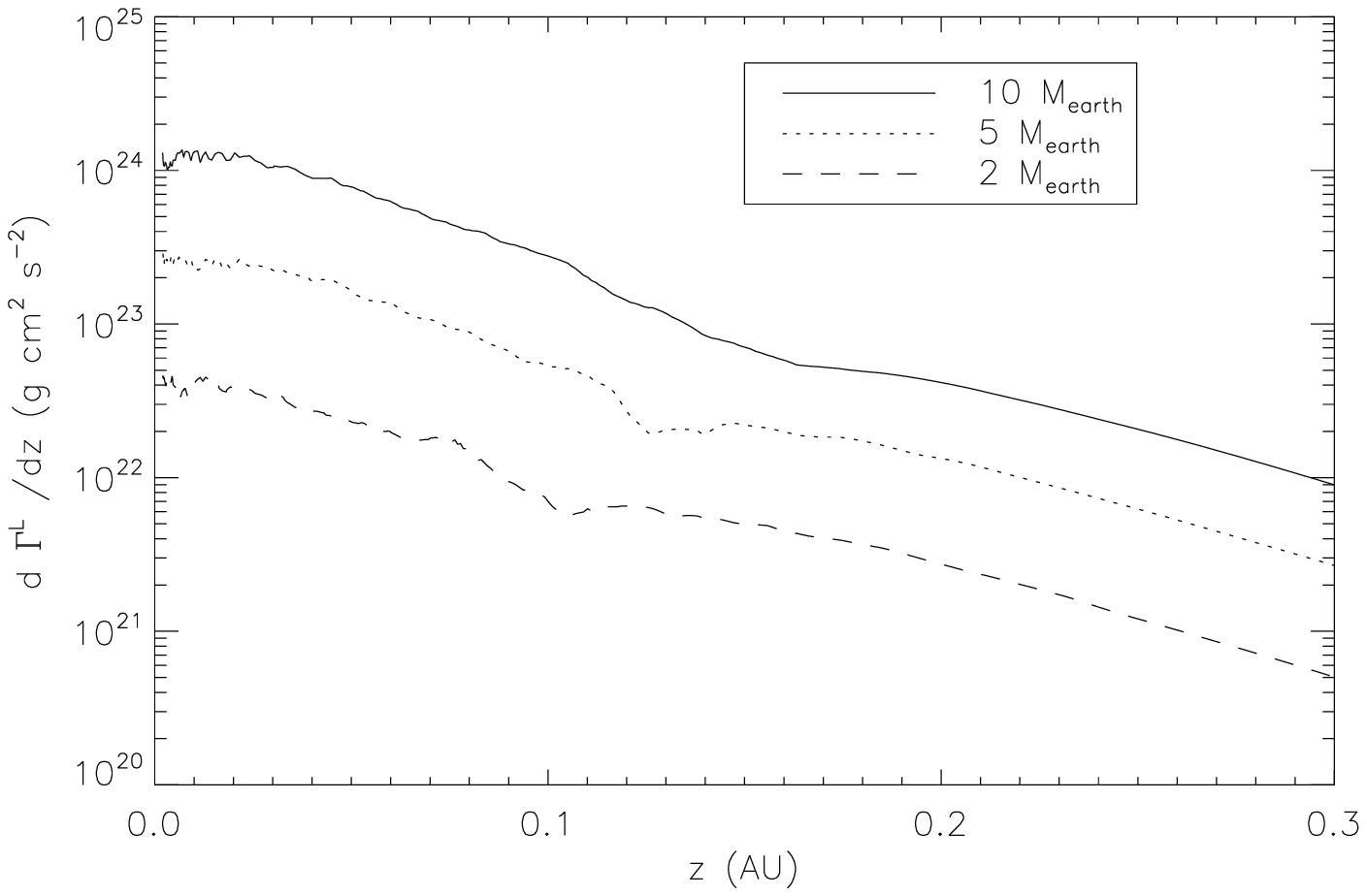}
\caption{Well mixed dust models. The torque exerted by planets on the disc, as a function of the distance from 
the mid-plane for an S07 model ($\Sigma \propto r^{-1}$). The solid line 
denotes the case of 10 $M_{\oplus}$, the dotted line is for 5 $M_{\oplus}$, the dashed line is for 
2 $M_{\oplus}$. Every planet is placed at 3 au. Every case shows the dips formed by the presence of the 
planet. The torque is an increasing function of planetary mass, as expected.}
\label{fig5}

\includegraphics[width=8.4cm]{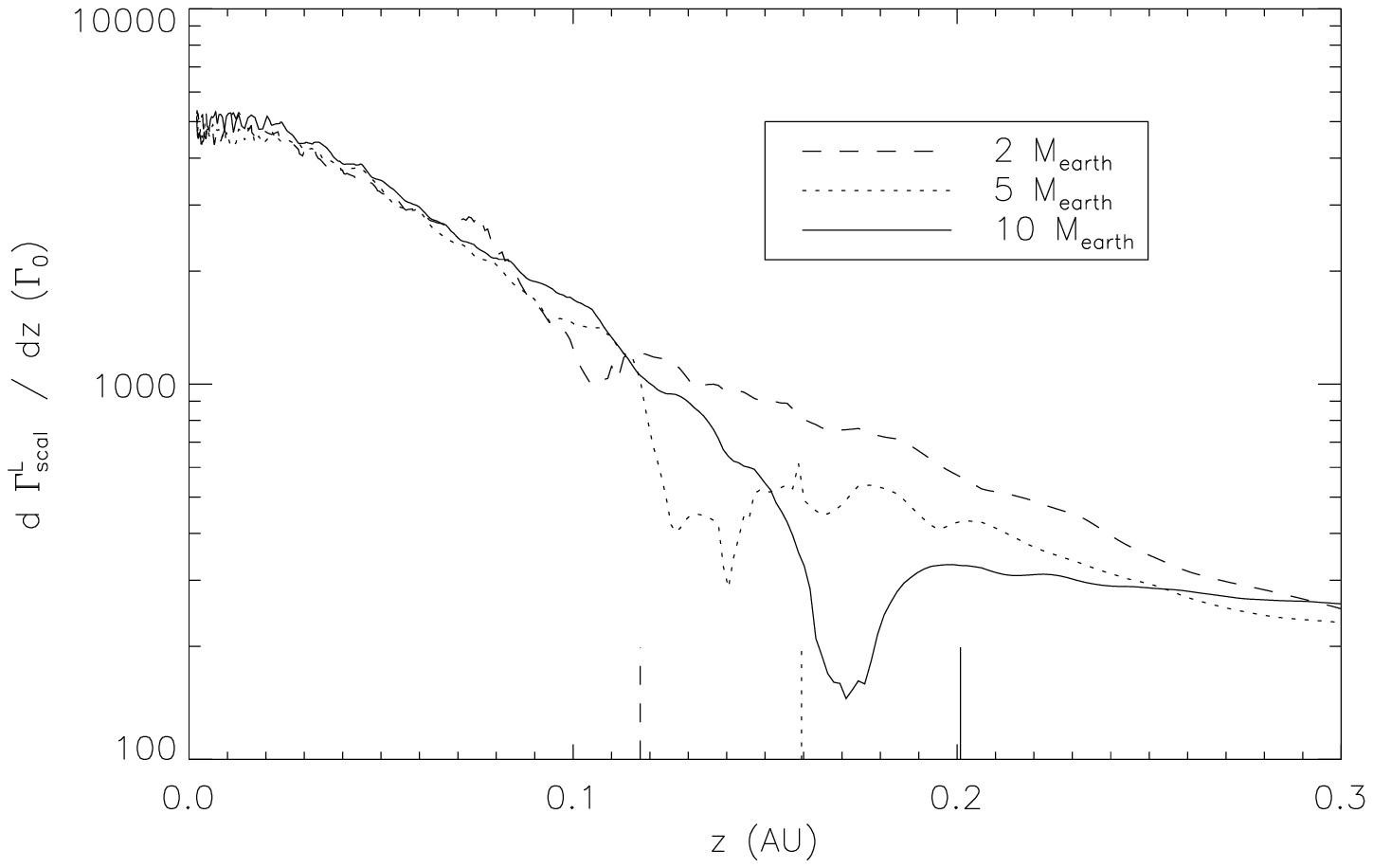}
\caption{Well mixed dust models. The scaled torque as a function of the distance from the mid-plane for 
an S07 model ($\Sigma \propto r^{-1}$). The solid line denotes the case of 10 $M_{\oplus}$, the dotted 
line is for 5 $M_{\oplus}$, the dashed line is for 2 $M_{\oplus}$. Every planet is placed at 3 au. The Hill 
radius for each planet is represented by the corresponding vertical line. The scaled torque takes a minimum 
value slightly below the Hill radius due to higher temperatures caused by the gravitational force of the 
planet.}
\label{fig6}
\end{center}
\end{figure}

\begin{figure}
\begin{center}
\includegraphics[width=8.4cm]{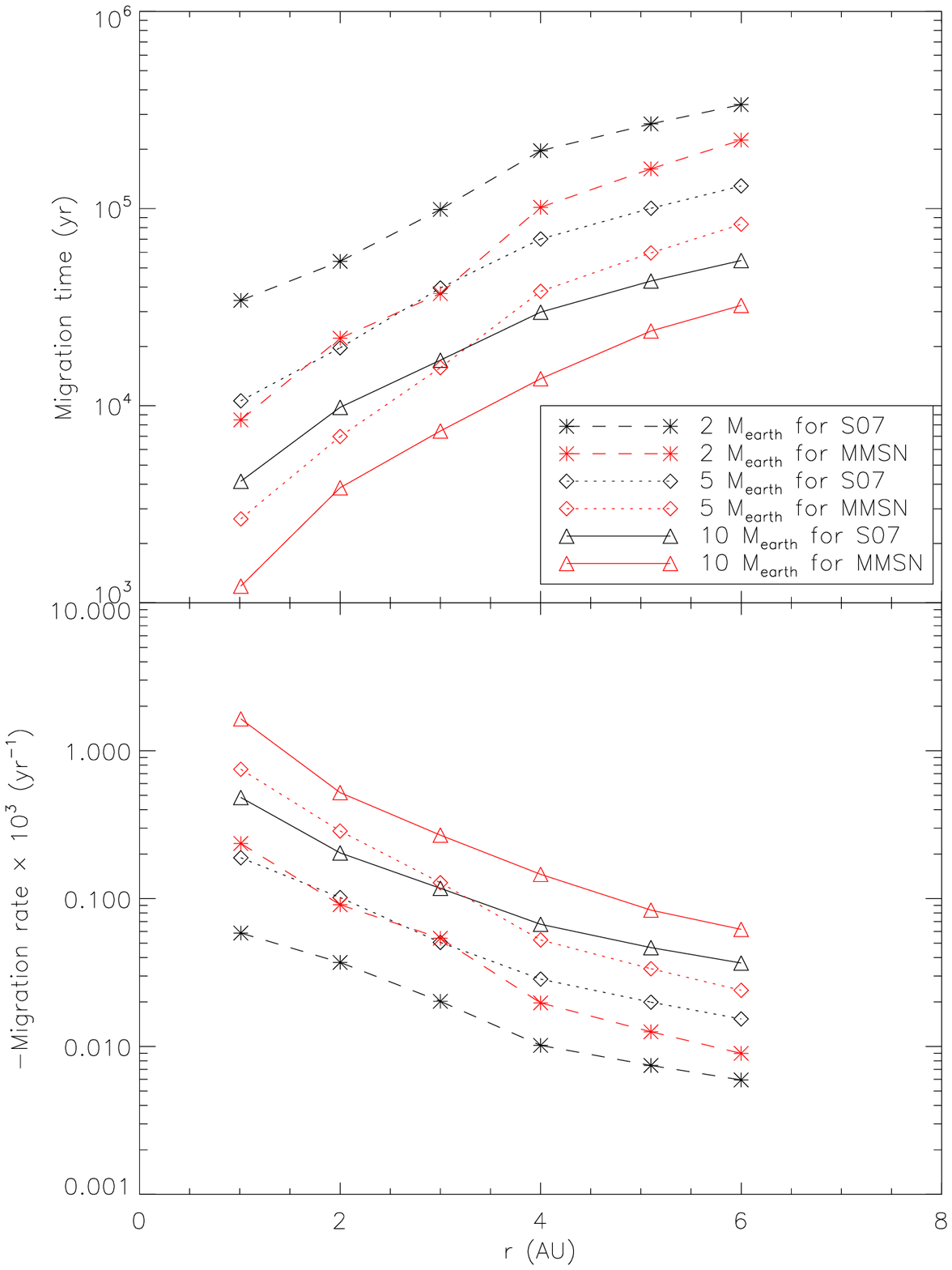}
\caption{Well mixed dust models. The migration time and rate as a function of orbital radius of a planet 
on the top and bottom panels, respectively. The solid line denotes the case of 10 $M_{\oplus}$, and the 
dotted line is for 5 $M_{\oplus}$, the dashed line is for 2 $M_{\oplus}$. The results of the MMSN disc 
models are denoted by the red lines while those of the S07 one are by the black lines. For both disc models, 
the migration time becomes smaller and the rate becomes larger as the planetary 
mass increases and the orbital radius decreases. The MMSN disc models produce more rapid migration.}
\label{fig7}

\includegraphics[width=8.4cm]{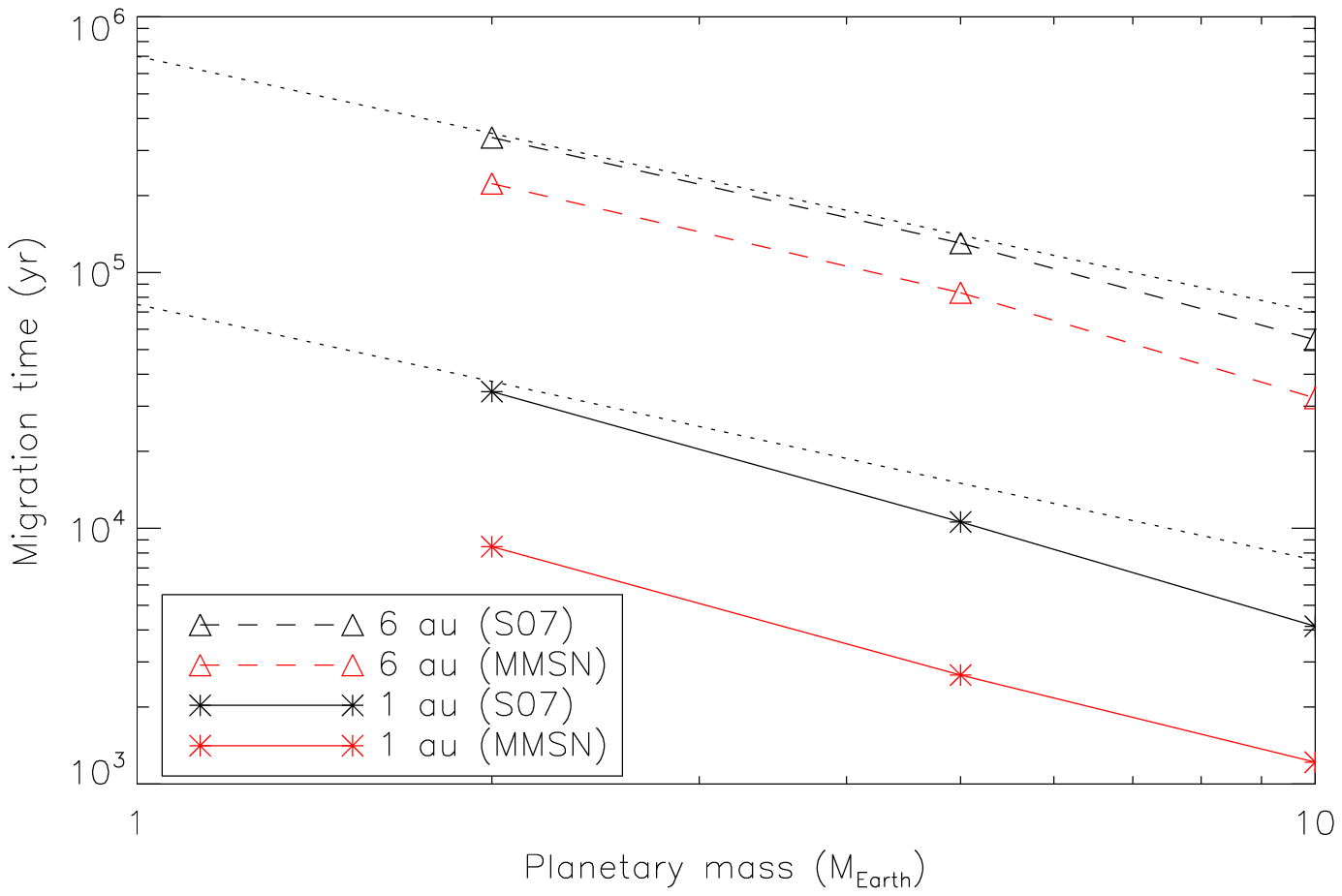}
\caption{Well mixed dust models. The migration time as a function of planetary mass. 
The solid lines with asterisks denote the case of $r_p=1$ au, the dashed lines with open triangles are 
for $r_p=6$ au. The results of the MMSN disc models are denoted by the red lines while 
those of the S07 one are by the black lines. The two dotted lines denote for a  
function $\propto 1/M_p$. For both disc models, the results deviate from this function due to the presence 
of the planet. The deviation increases with planetary masses. Also, our results approach the function as the 
orbital radius of the planet increases. This implies that the effects of the planet are diminished at 
larger disc radii due to the low disc density there.}
\label{fig8}
\end{center}
\end{figure}

\section{Results for the case of dust settling} \label{results2}

In this section, we examine the effects of dust settling in the presence of planets with various 
masses. We also discuss the difference between the well mixed and dust settling cases.  

\subsection{Torque}

We again focus on the behaviour of the torque, $ d \Gamma^L (r_p,z)/ dz$, exerted by a layer of 
gas at height $z$ (see equation (\ref{layer_torque})). As is the case of well mixed dust, the results of the two 
disc models are similar, so that we only present the results of the the S07 disc model ($\Sigma \propto r^{-1}$). 
We discuss the differences between them later.

\begin{figure}
\begin{center}
\includegraphics[width=8.4cm]{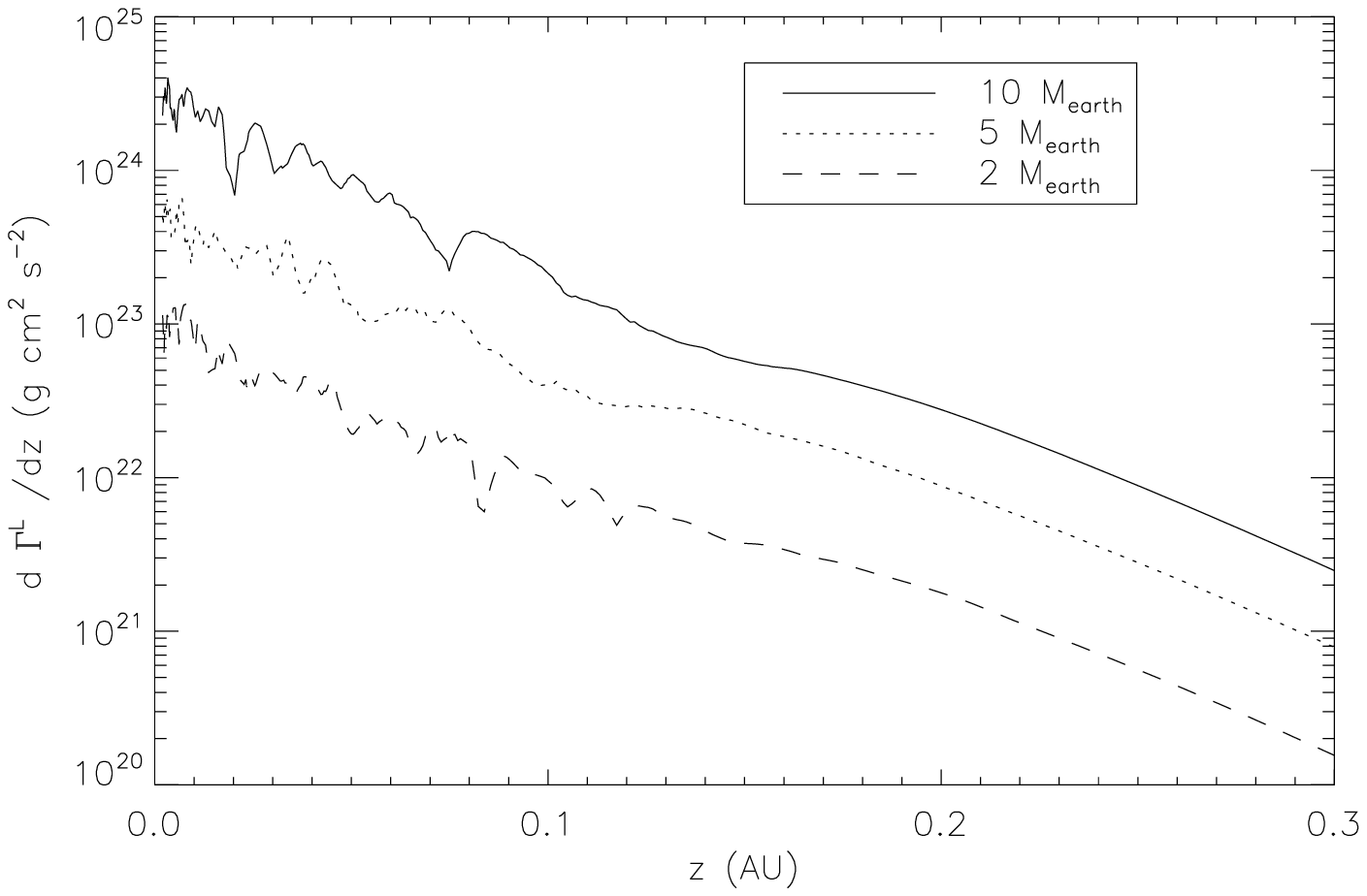}
\caption{Dust settling models. The torque exerted by a planet on the disc, as a function of the distance from the 
mid-plane for an S07 disc model ($\Sigma \propto r^{-1}$) (as Fig. \ref{fig5}). 
All three planets are placed at 3 au. As is the case of well mixed dust, all three lines show the dips 
formed by the presence of the planet and smoother behaviours. Compared with the case of no dust settling, the 
torque becomes stronger, especially around the mid-plane region and it decreases rapidly.}
\label{fig9}

\includegraphics[width=8.4cm]{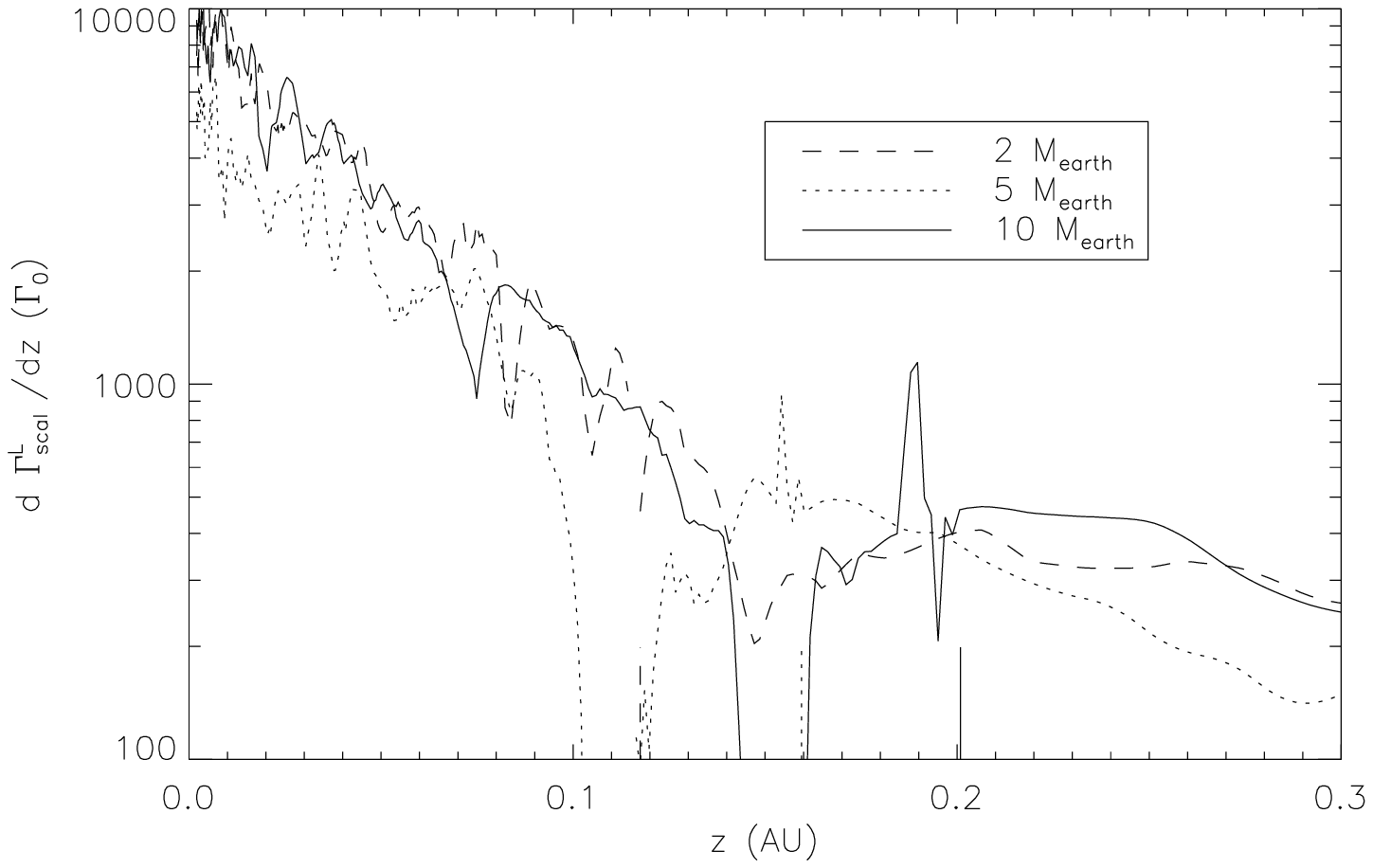}
\caption{Dust settling models. The scaled torque as a function of the distance from the mid-plane for an S07 disc 
model ($\Sigma \propto r^{-1}$) (as Fig. \ref{fig6}). All three planets are placed at 3 au. 
Dust settling increases the scaled torque, especially in the mid-plane region due to lower temperatures. 
In addition, it shifts the position of a minimum value closer to the mid-plane and enhances the dips 
formed by the presence of the planet, compared with the well mixed dust case (see Fig. \ref{fig6}).}
\label{fig10}
\end{center}
\end{figure}

Fig. \ref{fig9} shows the torque $d \Gamma^L (r_p,z)/ dz $ as a function of $z$ for the planets with various masses. 
Compared with the well mixed case (Fig. \ref{fig5}), dust settling provides a stronger torque around the mid-plane 
region, as expected (see $\S$ \ref{torque}). This arises from lower temperatures in the mid-plane region induced 
mainly by dust settling. Since a small change at such lower temperatures strongly affects the resonant positions, 
the distribution of the torque slightly fluctuates, especially around the mid-plane region. Although one might 
worry that such fluctuations cause spurious effects, we checked that the resultant migration time and rate converge 
in our grid system \citep{h08}. The common features with well mixed dust are that the tidal torque becomes stronger 
with increasing planetary mass, and that its distribution shows the effects of the planet (although they are small). 

When the MMSN disc models are used, the torque becomes stronger because of lower temperatures around the planet, 
as discussed in the well mixed case (also see the right column of Table \ref{table2}). Thus, dust settling 
accelerates the migration for both disc models by up to a factor of 1.8.  
(see Fig. \ref{fig11} and Table \ref{table3}). 

In order to disentangle the thermal effects on the torque, we again consider the scaled torque 
$d \Gamma^L_{scal}/ dz$ 
(see equation (\ref{scaled_torque}) and Fig. \ref{fig10}). Compared with the well mixed case, dust settling enhances 
the structure of the dips and moves them slightly closer to the mid-plane. This is because dust settling exaggerates 
the effects of the planets by transforming the disc from flared to flatter structures. Thus, dust settling 
diminishes contribution from the scaled torque around $z=r_H$ because of the higher temperatures. At the same time, 
it strengthens the scaled torque around the mid-plane region by dust settling as well as planetary gravity. As with 
the well mixed case, the increment of the torque around the mid-plane more than compensates the decrement around 
$z=r_H$, and consequently dust settling  more accelerates planetary migration (Table \ref{table3}).

Compared with the S07 disc model, the MMSN disc models tend to move the dips away from the mid-plane while 
the values around the mid-plane region slightly increase. These trends are the same as the well mixed case 
as already discussed. Compared with the well mixed case, however, the scaled torque $d \Gamma^L_{scal} / dz$ 
for the case of dust settling is more sensitive to the change in the background disc structure. This occurs 
because dust settling produces cooler and denser disc structures.   

\subsection{Migration time \& rate}
   
\begin{figure}
\begin{center}
\includegraphics[width=8.4cm]{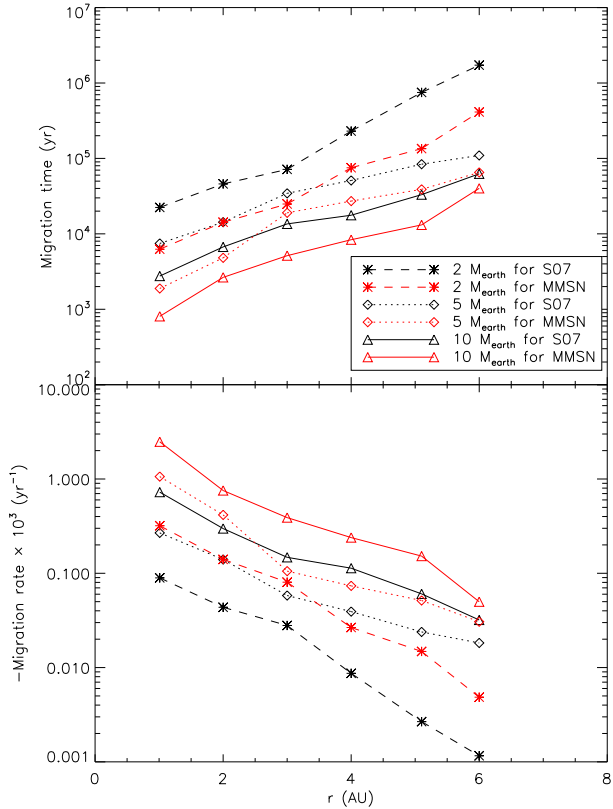}
\caption{Dust settling models. The migration time and rate as a function of orbital radius of a planet 
on the top and bottom panels, respectively (as Fig. \ref{fig7}). For both disc models, the migration time 
becomes smaller and the rate becomes larger due to dust settling. As with the case of well mixed dust, 
the MMSN disc models result in more rapid migration.}
\label{fig11}
\end{center}
\end{figure}

Fig. \ref{fig11} shows the migration time and rate as a function of orbital radius of a planet. As discussed 
above, dust settling leads to more rapid migration for both disc models. Table \ref{table3} summarises the 
difference of the migration time between the fully mixed and dust settling cases for both disc models at 1 au. 
Interestingly, the difference between the well mixed and dust settling cases is almost identical for both disc 
models. 

When the results of the two disc models (MMSN vs S07) are compared, the difference diminishes 
(see the right column of Table \ref{table2}). This arises from that dust settling gives similar temperature 
structures for the two disc models, as discussed in $\S$ \ref{disk_ds}. 

\begin{table}
\caption{The ratio of the migration time without to with dust settling at 1 au}
\label{table3}
\begin{tabular}{ccc}
\hline
                 & MMSN disc models & S07 disc models \\ \hline
2 $M_{\oplus}$   & 1.3 (1.5)        & 1.5 (1.5)       \\
5 $M_{\oplus}$   & 1.4 (1.5)        & 1.4 (1.4)       \\
10$M_{\oplus}$   & 1.5 (1.8)        & 1.5 (1.7)       \\
\hline
\end{tabular}

The maximum difference found in discs is given in the brackets
\end{table}

\subsection{Mass dependence}

In this subsection, we briefly discuss the combined effects of planets and dust settling on the torque, because 
their behaviors are the same as those for the well mixed case. We confirmed that the deviation from the 
function ($\propto 1/M_p$) becomes larger with increasing planetary mass, and hence planetary migration becomes 
faster. This implies that the distortion created by a planet is similar to dust settling even though the former 
is local and the latter is global. Both compress the disc density and lower the temperature in the 
mid-plane region. 

Also, we checked that the deviation generally becomes smaller with increasing the distance from the 
central star. This arises from the combined effects of dust settling and the power-law structure of the surface 
density. The flatter disc shape, (which is a result of dust settling,) reduces the grazing angle, resulting in 
a small decrement by the dip formation. At the same time, the power-law structure diminishes the effects of planets 
as discussed at the well mixed case. 

\section{Discussion} \label{discus}

\subsection{Other heat sources}

We have neglected viscous heating of discs in our models. We have assumed instead that stellar 
irradiation plays the main role in heating \citep{cg97}. As discussed in Paper I, viscous heating dominates 
stellar irradiation only within 0.1 au for M star systems while the turning point is shifted to a few au for 
classical T Tauri systems \citep[e.g.][]{js04}. Thus, viscous heating cannot affect our findings for M star 
discs beyond 0.1 au. 

In addition, we have neglected the accretion luminosity of planets. This heat source arises from the 
accretion of surrounding gas, dust, and planetesimals onto the planets driving the release of their extra 
gravitational energy. As discussed in Paper I, the resultant temperature profile roughly goes to $r^{-1/2}$ 
due to the inverse square law. This higher temperature in the vicinity of the planets can slow down their 
migration since the resonant positions are pushed away from them due to this higher pressure. In order to 
address the issue precisely, further detailed numerical studies are needed because only a realistic 
density distribution of circumplanetary discs allows one to accurately calculate the temperature. Also, 
such density and temperature structures affect corotation torques, which have been neglected in this study. 

\subsection{Neglect of corotation torque} \label{dis_ct}

Ever since \citet{pm06} numerically discovered an entropy-related corotation torque or horseshoe drag, which can 
be potentially larger than Lindblad one, the importance of corotation has been reexamined in the literature. 
However, recent progress has revealed that the entropy-related torque is not as important as 
expected (see Appendix \ref{app2} for more discussion). One of the reasons is that the adiabatic 
approximation, which establishes the entropy-related torque, readily breaks down in thermal equilibrium 
discs \citep{pbk10}. Consequently, the Lindblad torque likely controls planetary migration over larger 
parameter space than the corotation one. It is one of the reasons why corotation torque can be safely 
neglected in our disc models. One caveat is that our disc models cannot capture the horseshoe structure 
which is crucial for the horseshoe drag. Therefore, the proper treatment of this region is desired 
for accurately estimating the corotation torque. In addition, our axisymmetric assumption may affect 
the Linblad resonances as well particularly through the influence of the planet's gravitational field. 
We do not expect that our results are changed significantly.

\subsection{Synthesis of results}

In the above two sections, we examined the effects of planetary gravity and dust settling on the tidal torque in 
detail. In this subsection, we integrate our results and discuss how our results map to the literature.

\begin{figure}
\begin{center}
\includegraphics[width=8.4cm]{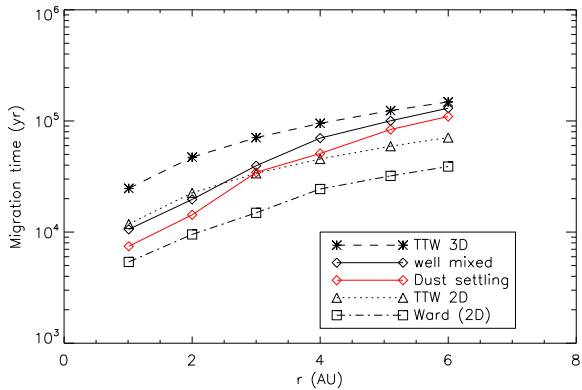}
\caption{The migration time as a function of the distance from the star. We adopt the S07 disc models 
with a 5 $M_{oplus}$ planet. Our results 
are denoted by the black and red solid lines for the well mixed and dust settling cases, respectively. For 
comparison purpose, the results of TTW02 for 3D discs are denoted by the black dashed line, those of TTW02 for 2D 
discs are by the black dotted line, and those of \citet{ward97} are by the black dash-dotted line. Our results 
show an transition in the migration time from the 2D to 3D calculations.}
\label{fig12}
\end{center}
\end{figure}

We proceed by comparing our results with those of \citet{ward97} and TTW02 (Fig. \ref{fig12}). We adopt 
the results for the S07 disc model with a 5 $M_{\oplus}$ planet. Fig. \ref{fig12} shows an interesting transition 
behaviour in the tidal torque for radiatively heated discs. For planets with small orbital radii 
($r\lesssim 3$ au), our results for the well mixed and dust settling cases are close to the results derived from 
2D discs, while, for planets with large orbital radii ($r \gtrsim 3$ au), ours are close to the ones derived from 
3D discs. Obviously, this arises from the complex combination of the effects of planets on the thermal and density 
structures of discs, as discussed above. Thus, in order to accurately estimate the migration time, it is essential 
to consider the effects of stellar irradiation and planetary gravity simultaneously.

For both cases, the migration time is shorter than the disc lifetime, which is a order of $10^6$ years. Indeed, 
dust settling aggravates the problem of type I migration. Therefore, it is clear that robust slowing mechanisms are 
needed to resolve the problem. HP10 and \citet{mpt09} demonstrated that dead zones act as barriers for it while 
\citet{il08v} considered that ice lines in layered structures play such a role. 

\subsection{Gap-opening mass \& type II migration} \label{typeII}

We have focused so far on type I migration. For completeness, we now examine the effects of dust settling on 
gap-opening mass planets which undergo type II migration in this section. 

Gap-opening mass is well discussed in the literature \citep[e.g.][references herein]{mp06}, so that we briefly 
present criteria above which planets can open up a gap . For inviscid discs, a gap-opening mass is
\begin{equation}
 \frac{M_p}{M_*} \geq 3 C \left( \frac{h_p}{r_p} \right)^{3}.
\end{equation}
We introduce a coefficient $C$ into the original condition, in order to take into account (some) possible 
reduction of the 
tidal torque. In many previous studies, the gap-opening criterion is derived under the 2D approximation for discs. 
TTW02, however, showed that the tidal torque in 3D is a factor of 2-3 smaller than that in 2D. Since the 
gap-opening mass involves the tidal torque, the mass for 3D discs can increase by at least a factor of 2 or 3. 
We therefore take $C=3$ here. We leave a more detail study on it to a future publication.

For viscous discs, a gap-opening mass is
\begin{equation}
 \frac{M_p}{M_*} \geq C \sqrt{40 \alpha \left( \frac{h_p}{r_p} \right)^{5}},
\end{equation}
where $\alpha$ is the strength of turbulence. The presence 
of viscous torque which plays a counteractive role in gap-opening, that is, it closes a gap.

\begin{figure}
\begin{center}
\includegraphics[width=8.4cm]{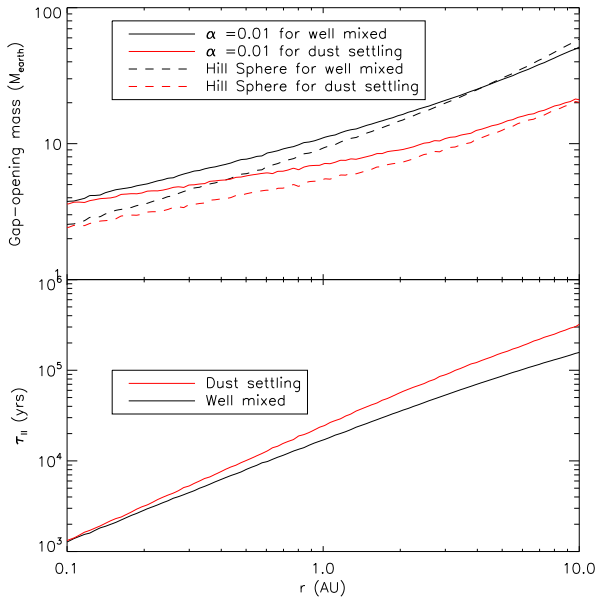}
\caption{The gap-opening mass and the timescale of type II migration as a function of the distance (on top and 
bottom panels, respectively). We adopt the S07 disc model without planets. The gap-opening masses for viscous 
discs are shown with solid lines while those for inviscid discs with dashed lines. For both panels, the results 
of the well mixed case are denoted by black lines while those of the dust settling case are by red lines. The 
gap-opening masses for the viscous discs are larger than those for the inviscid discs, as expected. Also, dust 
settling reduces the gap-opening mass, compared with the well mixed case. For the timescale of type II migration, 
dust settling results in slower migration than the well mixed case.}
\label{fig13}
\end{center}
\end{figure}

Fig. \ref{fig13} (Top) shows the gap-opening mass as a function of the distance from the star. In general, the mass 
derived from viscous discs is larger than that from inviscid discs. Both criteria for the dust settling case 
are smaller than those of the well mixed case, as expected. This is a consequence of having flatter discs which 
arise from dust settling. We note that, even for the dust settling case, all planets we have considered 
(2, 5, 10 $M_{\oplus}$) satisfy the criterion of type I migration (i.e. not open-up a gap) for $r=1$ au to 
$r= 10$ au. We found that dust settling reduces the gap-opening mass at 10 au, respectively, for inviscid discs by 
about a factor of 2.5, for viscous discs by about a factor of 2. 

In the above analysis, there is no time-dependency. \citet{pm06b} simulated the interaction of gas and dust 
in discs in the presence of massive planets, using a two-fluid approach. Although they cannot include dust 
settling, since they considered 2D discs, they demonstrated that dust-gaps are more readily formed than gas-gaps. 
Furthermore, they found that dust removed from the location of planets is accumulated at mean-motion resonances. 
Combined with such simulations, gap opening masses for dust settling are likely to be much lower.

Once a planet opens up a gap in the disc, it undergoes Type II migration \citep{ward97}. This type of migration is 
controlled by the viscous evolution of the gas in discs. The timescale is generally written as 
\begin{equation}
 \tau_{II}= \frac{1}{3 \alpha \Omega_p} \left( \frac{r_p}{h_p} \right)^2 (yrs),
\end{equation}
where $\Omega_p$ is the angular frequency at $r=r_p$ \citep{t03b}. Thus, type II migration is also affected 
by dust settling via the reduced disc scaleheight (the bottom panel of Fig. \ref{fig13}). We found that dust 
settling increases the timescale by about a factor of 2 at 10 au.

As a result, dust settling somewhat alleviates the problem of planetary migration in two respects: the 
reduction of the gap-opening mass and the slowing of type II migration. In discs with dust settling, low mass planets undergo 
faster type I migration. At the same time, such planets more readily open-up a gap, and switch to type II migration 
which is generally slower than type I. 

\section{Conclusions} \label{conc}

We have carefully examined Lindblad torques in radiatively heated discs. In order to make our disc models 
realistic, we have included dust settling and the gravitational force of planets. We have demonstrated that 
dust settling accelerates type I migration by up to a factor of about 2. This arises from the lower temperatures 
of the mid-plane region and the resultant flatter shape of discs. Both are the consequence of dust settling.

We have also investigated the mass dependence on the torque, by considering planets with various masses (2, 5, 10 
$M_{\oplus}$). We have found that the gravitational force of planets has two effects on the torque. 
One is the reduction of the torque contribution around $z=r_H$. 
The other is to lower the disc temperature in the mid-plane which increases the torque. In our disc models. the 
increment is dominant over the reduction, so that massive planets undergo even more rapid migration.

We have also considered disc models with two different surface density profiles: $\Sigma \propto r^{-1}$ 
(S07) and $\Sigma \propto r^{-3/2}$ (MMSN). We found that, for the well mixed case, planets for the MMSN models 
migrate faster than those for the SO7, by up to a factor of 4. For the dust settling case, the difference becomes 
small (about a factor of  2).

Finally, we have studied the effects of dust settling on the gap-opening mass and type II migration. The reduction 
of the scale height by dust settling decreases the gap-opening mass and increases the timescale of type II 
migration. Thus, dust settling may slightly reduce the problem of planetary migration for sufficiently massive 
planets. Our results show that dust settling has a clear and significant effect on planetary migration of both 
low and high mass planets. It is clear that an even more effective slowing mechanism is required in order to slow 
type I migration. Currently, two types of inhomogeneity in discs could do the job. HP10 and \citet{mpt09} 
demonstrated that dead zones provide two robust barriers while ice lines considered by \citet{il08v} provide 
another robust barrier. In our next paper, we will incorporate these barriers in one disc model, and discuss their 
roles in producing the mass-period relation.   

\section*{Acknowledgments}

Our simulations were carried out on computer clusters of the SHARCNET 
HPC Consortium at McMaster University. We thank an anonymous referee for useful comments on our 
manuscript. YH is supported by McMaster University, as well as by 
Graduate Fellowships from SHARCNET and the Canadian Astrobiology Training Program (CATP). REP is supported 
by a Discovery Grant from the Natural Sciences and Engineering Research Council (NSERC) of Canada.

\bibliographystyle{mn2e}

\bibliography{mn-jour,adsbibliography}

\appendix

\section{The Lindblad torque formula} \label{app1}

Here, we describe the formulation of the Lindblad torque. Since we calculate the torque density, 
the Lindblad resonances $r$ are incorporated into the wavenumber $m$;

\begin{equation}
m(r,z)=\sqrt{\frac{\kappa^2}{\left( \Omega-\Omega_p \right)^2 -c_s^2/r^2}},
\label{wave_num}
\end{equation}
where $\Omega_p=\sqrt{GM_*/r_p^3}$ is the angular velocity of a planet at the distance $r_p$ from the 
host star with its mass $M_*$, $c_s (\propto \sqrt{T})$ is the sound speed, $T$ is the disc temperature, 
the angular velocity $\Omega$ is 
\begin{equation}
\Omega^2(r,z)=\Omega_{Kep}^2{\left( 1 + \frac{z^2}{r^2}\right)^{-3/2}} 
              + \frac{1}{r\rho}\frac{\partial p}{\partial r},
\label{omega}
\end{equation}
$\Omega_{Kep}=\sqrt{GM_*/r^3}$ is the Keplerian frequency, $\rho$ is the gas density, $p=c_s^2\rho$ is the 
pressure, and the epicyclic frequency $\kappa$ is
\begin{equation}
\kappa^2(r,z)=\frac{1}{r^3}\frac{\partial}{\partial r} \left( r^4 \Omega^2 \right).
\end{equation}
Thus, $m$ becomes a function of $r$ that represents the resonant positions. Note that the vertical 
thickness of discs are included in equation (\ref{omega}).

For a planet with mass $M_p$, the Lindblad torque density $d^2 \Gamma^L/dzdr$ of a layer at the 
height from the mid-plane $z$ may be written as 
\begin{equation}
\frac{d}{dz} \left( \frac{d\Gamma^L}{dr}(r,z) \right) = \epsilon  
                                \frac{2\mu ^2 \rho r_p^4 \Omega_p^4}{r(1+4\xi ^2)\kappa^2}m^4\psi ^2,
\label{torque_density}
\end{equation} 
where 
\begin{equation}
\epsilon = 
\begin{cases}
 + &  \mbox{ for the outer resonances},  \\ 
 - &  \mbox{ for the inner resonances}, 
\end{cases}
\label{epsilon}
\end{equation}
$\mu=M_p/M{*}$, $\xi = mc_s/r\kappa $, the forcing function $\psi$ is
\begin{equation}
\psi(\alpha_r,\zeta)=\frac{\pi }{2}
                     \left[ -\varepsilon \frac{\alpha_r}{m} \frac{db^m_{1/2}(\alpha_r,\zeta )}{d \alpha_r} 
                      +2\sqrt{1+\xi ^2}b^m_{1/2}(\alpha_r,\zeta )\right],
\label{ff}
\end{equation}
and the Laplace coefficient $b^m_{1/2}$ is
\begin{equation}
b^m_{1/2}(\alpha _r,\zeta)=\frac{2}{\pi}
                            \int^{\pi}_{0}\frac{\cos m \theta d\theta }
                                 {\sqrt{1-2\alpha _r\cos \theta +\alpha_r ^2+\zeta ^2}},
\label{lc}
\end{equation}
where $\alpha _r = r/r_p$ is the normalised Lindblad resonant position and $\zeta =z/r$ diminishes 
the $b^m_{1/2}$ due to the vertical thickness of discs. Note that the integration of equation 
(\ref{torque_density}) in terms of $z$ with $\zeta=0$ leads to the famous analytical formulae of the Lindblad 
torque density in 2D discs \citep[see][MG04]{ward97}. Also, note that the tidal torque shown above is 
exerted by the planets on discs. Thus, the positive sign of $\epsilon$ in equation (\ref{epsilon}) means that 
the planets exert the tidal torques on discs, resulting in the loss of angular momentum from the planets and 
vice versa.

The tidal torque reaches its maximum value at the resonant position which is located at about $2h/3$ from the 
planets. It decreases far from this point because the distance between the planets and the resonances is 
increased \citep{ward97}. Furthermore, it is negligible within the region between $r_p \pm 2h/3$ due to the 
gas pressure. In other words, the resonant positions are pushed away from the planets due to the effect of the 
azimuthal pressure, - the so-called pressure buffer \citep{a93}. This relation is clearly seen in equation 
(\ref{wave_num}). When $\Omega\approx \Omega_p$, $m$ becomes imaginary, meaning that there is no $m$ 
corresponding to such resonant position $r$. Thus, the dominant contribution to the total torque arises 
from the region with $r \simeq r_p \pm 2h/3$. 

This allows one to approximately replace $\cos \theta$ in equation (\ref{lc}) with $1-\theta^2/2$ 
\citep[GT80;][JS05]{ward97}. The Laplace coefficient $b^m_{1/2}$ then becomes
\begin{equation}
b^m_{1/2}(\alpha _r,\zeta) \approx \frac{2}{\pi\sqrt{\alpha_r}}K_0(\Lambda ),
\end{equation}
where $K_i$ is the modified Bessel function of the second kind of order $i$ and the argument $\Lambda$ is
\begin{equation}
\Lambda(\alpha_r,\zeta) =m\sqrt{\frac{(\alpha_r -1)^2+\zeta ^2}{\alpha_r}}.
\end{equation}
Substituting in this approximation and using $K_0'=-K_1$ \citep[e.g.][]{as72}, the forcing function $\psi$ 
becomes
\begin{equation}
\psi \approx \frac{\epsilon (\alpha_r^2-\zeta ^2-1)}{2\alpha_r\sqrt{(\alpha_r-1)^2+\zeta ^2}}K_1(\Lambda )
      + \left( \frac{\epsilon }{2m}+2\sqrt{1+\xi^2} \right) \frac{K_0(\Lambda )}{\sqrt{\alpha_r}}.
\end{equation}

Finally, the total torque exerted by the planets on discs is calculated by
\begin{equation}
\Gamma^L=\int^{\infty}_{-\infty}dz \int^{\infty}_{0}dr \frac{d}{dz} \left( \frac{d\Gamma^L}{dr}(r,z) \right).
\label{torque_tot}
\end{equation}

\section{The problems with corotation torque} \label{app2}

Here, we discuss corotation torque and summarise its problems. The corotation torque was originally 
derived by GT80. The corotation resonances arise when the orbital frequency of the gas $\Omega$ is 
identical to the orbital frequency of the planet $\Omega_p$.  The analytical formula is 
\begin{equation}
\Gamma_m^C=\frac{m\pi^2}{2}\left[ \left( \frac{d\Omega}{dr}\right)^{-1}
                                    \frac{d}{dr}\left( \frac{\Sigma}{B}\right)\phi_m^2 \right],
\label{corotation_torque}
\end{equation}
where
\begin{equation}
B=\frac{1}{2r}\frac{d}{dr}\left( r^2 \Omega \right)
\end{equation}
is one of the Oort constants or the vorticity of gas, and 
\begin{equation}
\phi_m=-\frac{GM_p}{r_p}b^m_{1/2}(\alpha _r)
\end{equation}  
is the $m$th order Fourier component of the potential of the planets. Thus, the sign of the corotation 
torque is determined by the radial gradient of vorticity per unit surface density, called the vortensity. For 
the Keplerian discs, $B\propto \Omega_{Kep}$. Consequently, $\Gamma_m^C$ vanishes with 
$\Sigma \propto r^{-3/2}$. Unfortunately, this first derived analytical formula had a discrepancy with 
numerical calculations \citep{kp93}. Furthermore, the mechanism of the exchange of angular momentum between 
the planets and the gaseous discs and its relation with the the gradient of the vortensity is physically 
not clear. 

The discrepancy between the analytical and numerical torques was resolved by TTW02 by replacing $\phi_m^2$ 
with $|\phi_m+\eta _m|^2$, where $\eta_m$ is the $m$th Fourier component of the enthalpy perturbation. 
This implies that the gas pressure or temperature around the planet is crucial for a proper prescription of 
the corotation as well as the Lindblad torques since the enthalpy is defined by the pressure divided by 
density. The revised full analytical formula of the corotation torque has recently been derived by 
\citet{zl06}. We call the torque discussed above the linear corotation torque since it is derived from the 
linearized equations.

\citet{ward91} derived the corotation torque based on a fundamentally different approach. This corotation 
torque is often called a horseshoe drag to distinguish with the linear one. The horseshoe drag arises from 
the interaction between a planet and the fluid element moving in the vicinity of it. When the fluid element 
encounters the planet, the orbital radius of the element is shifted from the inner to the outer one 
relative to the planetary orbit. During this non-linear process, the fluid element gains 
angular momentum from the planet and vice versa since angular momentum of the disc increases with distance 
from the central star. When the orbital radius of the fluid element is shifted from the outer to the inner 
orbits, the situation is the opposite. As seen in numerical simulations \citep[e.g.][]{m01}, these shifts 
result in a horseshoe orbit. That is why it is called the horseshoe drag. Furthermore, adiabatic invariance 
forces the fluid to continue its circular motion \citep{ward91}, which prohibits the excitation of 
density waves and possibly also produces saturation (that is, the horseshoe drag becomes negligible). 

It is very interesting that the horseshoe drag also depends on the radial gradient of the vortensity which 
is explained by mapping process of the area occupied by the fluid element. Note that there is no reason to 
identify the linear corotation torque with the horseshoe drag since they are derived from different 
approaches. The formula of the horseshoe drag is 
\begin{equation}
\Gamma_m^{HS}= 4 \Sigma |A| B x_s^4 \left( \frac{d \ln (\Sigma/B)}{d \ln r}\right),
\label{horseshoe_drag}
\end{equation}
where $x_s$ is the half width of the horseshoe region and
\begin{equation}
A=\frac{1}{2} r \frac{d \Omega}{dr}
\end{equation}
is one of the Oort constants \citep{ward91}. Thus, the horseshoe drag strongly depends on $x_s$ although 
$x_s$ is treated as a parameter in any numerical simulations \citep{mdk06,pp09}.

Recently, \citet{pp09a} have reexamined the linear corotation torque and horseshoe drag and found that the 
linear one is only valid for the early stage of numerical simulations or for discs with very high viscosity 
($\alpha \sim 0.1$). 
Except for these, the total torque is well-represented by the sum of the Lindblad torque and the horseshoe 
drag. With the proper choice of $x_s$, the horseshoe drag is about 2-3 times larger (in magnitude) than the 
linear corotation torque, but it is still smaller than the Lindblad one for discs with $s<1$, where 
$\Sigma \propto r^s$. A more complete treatment for the horseshoe drag including the effects of 
the pressure is done by \citet{cm09} and confirms the same results.

In the above studies, the temperature which may be a function of radius from the star is assumed to be fixed 
during mass transfer. In other words, discs are assumed to be barotropic. In this case, the magnitude of the 
linear corotation torque or horseshoe drag is always smaller than the Lindblad one \citep{kp93}. Therefore, 
one can safely assume both torques to be negligible even if it is not saturated. The situation, however, 
can change by relaxing the isothermal assumption \citep{pm06}. For discs in which the energy conservation 
is fully considered, the linear corotation torque and horseshoe drag are determined not only by the radial 
gradient of the vortensity but also by the radial gradient of entropy \citep{bm08,pp08}. It is interesting 
that the entropy-related corotation torque, especially horseshoe drag, can dominate over the Lindblad one 
and control migration if discs have a large entropy gradient and are considered to be adiabatic. 
In the region with a large (in magnitude), negative entropy gradient, planets migrate outwards 
due to the large, positive horseshoe drag. The complete analytical formulae in 2D discs are derived by 
\citet{pbck09}, assuming saturation does not occur yet.

\citet{mc09} reexamined the corotation torque in adiabatic discs, taking into account the effects 
of pressure. They showed that the entropy-related torque does not arise from the overdense and underdense 
regions, as had been previously thought. Instead, they showed that entropy gradients make the horseshoe region 
considerably asymmetric, and this asymmetry exerts the entropy-related torque by the excitation of evanescent 
waves at the horseshoe sparatrices. Since these evanescent waves are excited by pressure disturbance that are 
a result of horseshoe orbits, the magnitude of the entropy-related torque depends on the perturbed pressure. 
This gives an explanation of the results of \citet{bm08} that the corotation torque in a very cold, adiabatic 
disc is identical to that in an isothermal disc.

Thus, the corotation torque or horseshoe drag can be important only for discs assumed to be adiabatic with high 
temperatures, that is, the radiative cooling timescale in discs is longer than the planet's orbital period. 
Otherwise, the direction of migration is determined by the Lindblad torque. The horseshoe drag, however, has 
some problems. One is the width of the horseshoe region, and the other is saturation. The former is 
characterised by the mass of the planets, but it is parameterized in numerical simulations partly because 
a planet is treated as a point mass object \citep{dkh03,pbck09} and partly because any simulation cannot 
infinitely resolve the horseshoe region \citep{mdk06,pp09}. For the latter problem, the horseshoe drag can 
be saturated (i.e. negligible) due to adiabatic invariance. 

We emphasise that the problem of saturation is essentially how the disc viscosity connects the horseshoe region 
to the rest of disc by transferring angular momentum between them \citep[e.g.][]{mc10}. This is crucial for 
understanding horseshoe drags. As a example, we summarise the work done by \citet{pbk10} who considered the 
saturation of the vortensity-related and entropy-related horseshoe drags. For the vortensity-related one, 
the saturation is controlled only by viscous diffusion \citep[also see][]{m01,m02,ward07}. For the discs 
with $\alpha \sim 10^{-5}$, the horseshoe drag becomes almost zero while it is unsaturated for the disc with 
$\alpha \sim 0.01$ (see their fig. 2). For the entropy-related one, which is more important for migration, 
the saturation is controlled by viscous as well as the thermal diffusion \citep{pp08}. It is 
important that \citet{pbk10} found the thermal diffusion is always much larger than the viscous diffusion in 
thermal equilibrium discs in which viscous heating is balanced with radiative losses. Since thermal 
diffusion defines the validity of the adiabatic approximation of discs, this finding strongly 
restricts the applicability of the entropy-related horseshoe drag. Indeed, they have confirmed that the 
entropy-related horseshoe drag is reduced by about a factor of 2 due to thermal diffusion 
(see their fig. 7). Although their findings are potentially changed by the mass of a planet which controls 
saturation \citep{pbk10}, we can conclude that the entropy-related horseshoe drag is not so 
important for migration as expected before. This is because high viscosity ($\alpha \sim 0.01$) is required 
for horseshoe drag to be unsaturated while it also boosts thermal diffusion, resulting in the discs 
considered to be (locally) isothermal. Consequently, only the vortensity-related horseshoe drag matters. 
For low viscosity ($\alpha \sim 10^{-5}$), any torque arising at the corotation resonances readily suffers 
from saturation. Therefore, it is reasonable that corotation torque is assumed to be negligible.  
     
The bottom line appears to be that the corotation torques or horseshoe drags are unlikely to play a dominant 
role in planetary migration.  

\bsp

\label{lastpage}

\end{document}